\RequirePackage{ifpdf}
\documentclass[11pt,a4paper]{article}
\usepackage{jheppubme}
\usepackage{etoolbox}
\usepackage{comment}
\usepackage{amsmath,amssymb,epsfig}
\usepackage{amsfonts}
\usepackage{verbatim}
\usepackage{latexsym}
\usepackage{amsthm}
\usepackage{amsbsy}
\usepackage{xspace}

\def\one{{\,\hbox{1\kern-.8mm l}}}
\newcommand{\Dslash}{\not{\hbox{\kern-4pt $D$}}}
\newcommand{\pdslash}{\not{\hbox{\kern-2pt $\partial$}}}
 
\newcommand{\cN}{\mathcal{N}} 
\newcommand{\cM}{\mathcal{M}} \newcommand{\cR}{\mathcal{R}}

\newcommand{\Comment}[1]{{}}

\def\IZ{{\mathbb Z}}

\def\IR{{\mathbb R}}

\newcommand{\bc}{\begin{center}}
\newcommand{\ec}{\end{center}}
\newcommand{\ba}{\begin{array}}
\newcommand{\ea}{\end{array}}
\newcommand{\beq}{\begin{equation}}
\newcommand{\eeq}{\end{equation}}
\newcommand{\bea}{\begin{eqnarray}}
\newcommand{\eea}{\end{eqnarray}}
\newcommand{\bmx}{\begin{pmatrix}}
\newcommand{\emx}{\end{pmatrix}}

\newcommand{\be}{\begin{equation}}
\newcommand{\ee}{\end{equation}}

\newcommand{\del}{\partial}
\newcommand{\half}{{\frac{1}{2}\,}}

\newcommand{\tPsi}{{\tilde\Psi}}

\newcommand{\tchi}{{\tilde \chi}}

\newcommand{\eref}[1]{Eq.\,(\ref{#1})}

\newcommand{\zbar}{{\bar z}}
\newcommand{\abar}{{\bar a}}
\newcommand{\bbar}{{\bar b}}
\newcommand{\fbar}{{\bar f}}

\newcommand{\Dbar}{{\bar D}}

\newcommand{\phibar}{{\bar \phi}}
\newcommand{\psibar}{{\bar \psi}}

\newcommand{\vlambda}{{\vec\lambda}}
\newcommand{\vrho}{{\vec\rho}}

\newcommand{\cG}{{\cal G}}

\def\IB{\relax{\rm I\kern-.18em B}}
\def\IC{{\relax\hbox{\kern.3em{\cmss I}$\kern-.4em{\rm C}$}}}
\def\ID{\relax{\rm I\kern-.18em D}}
\def\IE{\relax{\rm I\kern-.18em E}}
\def\IF{\relax{\rm I\kern-.18em F}}
\def\II{\relax{\rm I\kern-.18em I}}
\def\IZ{\relax{\sf Z\kern-.35em Z}}
\def\Id{\relax{1\kern-.32em 1}}
\def\IG{\relax\hbox{$\inbar\kern-.3em{\rm G}$}}
\def\IR{\relax{\rm I\kern-.18em R}}
\newcommand\sfrac[2]{{\textstyle\frac{#1}{#2}}}

\newcommand\shalf{{\textstyle\frac12}}

\normalsize

\title{Universal RCFT Correlators from the Holomorphic Bootstrap}

\author{Sunil Mukhi$^a$}
\author{and Girish Muralidhara$^{a,b}$}

\affiliation{${}^a$Indian Institute of Science Education and Research,\\
Homi Bhabha Rd, Pashan, Pune 411 008, India}

\affiliation{${}^b$International Centre for Theoretical Sciences,\\
Shivakote, Hesaraghatta Hobli, Bengaluru 560 089, India}

\emailAdd{sunil.mukhi@gmail.com}
\emailAdd{girish.lm20@gmail.com}

\abstract{We elaborate and extend the method of Wronskian differential equations for conformal blocks to compute four-point correlation functions on the plane for classes of primary fields in rational (and possibly more general) conformal field theories. This approach leads to universal differential equations for families of CFT's and provides a very simple re-derivation of the BPZ results for the degenerate fields $\phi_{1,2}$ and $\phi_{2,1}$ in the $c<1$ minimal models. We apply this technique to compute correlators for the WZW models corresponding to the Deligne-Cvitanovi\'c exceptional series of Lie algebras. The application turns out to be subtle in certain cases where there are multiple decoupled primaries. The power of this approach is demonstrated by applying it to compute four-point functions for the Baby Monster CFT, which does not belong to any minimal series.}

\preprint{}


\begin{document}

\maketitle

\section{Introduction}

The conformal bootstrap in two dimensions was described in the pioneering work of Belavin, Polyakov and Zamolodchikov \cite{Belavin:1984vu} who discovered a series of minimal conformal field theories with central charge $c<1$. These theories are ``rational'' in the sense that they have a finite number of conformal primary operators. This approach was subsequently extended to RCFT's with Kac-Moody algebras (WZW models) and other extended symmetries including superconformal algebras, W-algebras and parafermion algebras (for extensive references, see \cite{DiFrancesco:1997nk}). In every case, the theories in question have finitely many primaries of the extended algebra (not just the Virasoro algebra). Their central charge, though greater than 1, falls into discrete series much like the minimal models. The universal concept behind solving for the correlation functions and partition functions of these theories is to implement all the constraints  arising from the symmetry algebras and then use the decoupling of special states called ``null vectors''. Finally, the coset construction seemed to complete the picture by expressing large families of theories as the coset of one WZW model by another. 

In recent years, interest has somewhat shifted to CFT's that are not rational (i.e. do not have a finite number of primary fields under any extended algebra), largely because of their application in the AdS/CFT correspondence. However, rational CFT's are far from being classified and there are several simple ones that do not fall into the categories listed above. In particular, in \cite{Hampapura:2015cea,Gaberdiel:2016zke,Hampapura:2016mmz} it has been shown that there are RCFT's with as few as two characters (thus, with one nontrivial primary upto complex conjugation and possible degeneracy) that are not minimal models of any kind. They are also not cosets of two WZW models, rather they are cosets of a meromorphic CFT of the type classified in \cite{Schellekens:1992db} by a WZW model. For such theories the spectrum of primaries, their fusion rules and the characters and partition function are all known. However there does not seem to be a straightforward way to identify their extended chiral algebra and null vectors, and thereby compute their correlators. 

The work of \cite{Hampapura:2015cea,Gaberdiel:2016zke,Hampapura:2016mmz} was based on an approach to the classification and computation of RCFT characters via modular-invariant holomorphic differential equations \cite{Mathur:1988na,Mathur:1988rx,Mathur:1988gt} (for related work, see also \cite{Anderson:1987ge,Eguchi:1988wh,Verlinde:1988sn,Naculich:1988xv,Bantay:2010uy}). We will refer to this approach as the ``Wronskian method''. In particular it was pointed out that given the method for computing characters, one can compute correlators in the same way by simply replacing modular invariance by crossing symmetry and characters by conformal blocks. In this way, for a few specific RCFT's, \cite{Mathur:1988rx,Mathur:1988gt} found crossing-invariant differential equations for their conformal blocks which turn out to be the very same equations found in \cite{Knizhnik:1984nr} from the null-vector approach.

In the present work we extend and systematise this approach and apply it to a variety of rational CFT's by computing four-point functions of their primaries. The only limitation is that the correlator should have two conformal blocks. In this way we precisely reproduce the differential equations found in \cite{Belavin:1984vu} for the Ising model and in fact for suitable primaries of all minimal models, making no use of null vectors. Next we apply it to the special WZW models of small rank and level 1 first classified as a series of two-character CFT's in \cite{Mathur:1988na}. Following the recent literature, we describe this set as the Deligne-Cvitanovi\'c exceptional series (see for example \cite{Beem:2017ooy} and references therein). We determine the conformal blocks simply by plugging in the CFT data into our master formula. This could not have been done in the traditional approach, since the null vector structure depends sensitively on the chiral algebra and differs from one series of models to another.

We will encounter some subtleties arising from two facts. One is that few-character CFT's tend to have ``spurious'' primaries, which would be present for representation-theoretic reasons but decouple in correlators. The other is related: sometimes the representation that is forced to flow in a conformal block differs from that of the primary of that block, as a result of which the leading singularity is determined by a secondary. Neither of these issues arises in the study of $c<1$ minimal models. The latter problem was identified in two specific WZW models in \cite{Mathur:1988rx,Mathur:1988gt} and in some more general situations in \cite{Fuchs:1989pp}. Here we investigate it in a broader context and attempt to formulate a general picture of how it affects the analysis.

As an example of a non-minimal model (of any extended chiral algebra) we consider the Baby Monster CFT, \cite{Hoehn:thesis,Hoehn:Baby8} which was recently described as a generalised coset of the Monster CFT in \cite{Hampapura:2016mmz}. This three-character RCFT has the same fusion rules as the Ising model, though the primaries have a high degree of degeneracy corresponding to representations of the Baby Monster group. In this case too, we are able to compute the four-point correlation functions of one of the primaries. We leave a more detailed discussion of coset and coset-like theories, including the Baby Monster, for the future.

The plan of this paper is as follows. In Section \ref{Wronskian-single} we review the Wronskian differential equation approach to conformal blocks in the simplest case where the primaries do not have any degeneracy, and apply the result to reproduce some of the famous equations of \cite{Belavin:1984vu}. In Section \ref{Wronskian-multiple} we generalise this method to the case where primaries have degeneracies, considering separately the case of real and complex primaries. We then apply these results to the series of two-character CFT's classified in \cite{Mathur:1988na}. In Section \ref{babymon} we apply the same procedure to the Baby Monster CFT. The computation of monodromies of conformal blocks, normalisation factors, and some explicit expressions for the blocks in the Baby Monster case, are relegated to a set of Appendices.

\section{Wronskian method for RCFT correlators: single-component primaries}
\label{Wronskian-single}

\subsection{Differential Equation for Conformal Blocks}

In this section we explain the Wronskian method to obtain differential equations for correlation functions of primaries in RCFT's. We do this here for single-component primaries, i.e. the case of no degeneracy, and will consider multi-component primaries in the next Section. This is effective not just for theories having a small number of primary fields, but for any RCFT where the four-point correlator is made up of two (or less) conformal blocks. A similar method has been used to classify characters, and hence partition functions, in such RCFT's \cite{Mathur:1988na,Mathur:1988rx,Naculich:1988xv,Hampapura:2015cea}. In situations where the RCFT correlator is already known to satisfy a differential equation arising from null vectors, this method reproduces the same equations. However it is more powerful in that it generalises to RCFT's where the null vector structure is not well-known, and can be used to extract both the correlators and characters of such theories.

The general theory  is as follows. Suppose we consider a four-point correlation function:
\be
\cG(z_i,\zbar_i)=\langle \phi(z_1,\zbar_1) \phi(z_2,\zbar_2) \phi(z_3,\zbar_3) \phi(z_4,\zbar_4)\rangle
\ee
where we consider all identical primary fields $\phi$ and assume they are real and have no degeneracy. In an RCFT, this correlation function is expressed as a finite sum over the square of locally holomorphic conformal blocks $f_1(z_i),f_2(z_i),\cdots,f_n(z_i)$:
\be
\cG(z_i,\zbar_i)=\sum_{\alpha,\beta=1}^n m_{\alpha\beta}f_\alpha(z_i)\fbar_\beta(\zbar_i)
\label{curlygf}
\ee
where $m_{\alpha\beta}$ are some constants. In this paper we will always consider $m_{\alpha\beta}=\delta_{\alpha\beta}$, i.e. left-right symmetric primaries, as well as $n=2$, i.e. two conformal blocks. Under crossing transformations, the blocks $f_\alpha$ transform into each other, but the correlation function is crossing-symmetric. We can also consider monodromies made up from repeated crossings, for example allow any one of the $z_i$ to encircle another one. Again the blocks $f_\alpha$ transform into each other and the full correlation function is invariant under all such monodromies.

Notice that we have on the one hand, non-holomorphic objects (correlation functions) that are invariant under crossing, and on the other hand holomorphic objects (the blocks) that are not invariant but transform into each other under crossing operations. But there is a very useful class of objects, the Wronskians of the blocks, that are simultaneously holomorphic {\em and} crossing-symmetric. They lead to a complex differential equation of which the conformal blocks are the independent solutions. This formulation turns out to be extremely powerful, as one can exploit holomorphy and crossing symmetry together. 

One constructs the Wronskians $W_k$ from the blocks $f_\alpha$ as follows. Pick one of the arguments of the correlator, say $z_1$, and use the derivative $\del_1=\frac{\del}{\del z_1}$. Here and in what follows, we explicitly exhibit the dependence of all functions on $z_1$, but should keep in mind that they also depend on $z_2,z_3,z_4$. Later, the latter dependence will be eliminated by a standard choice for $z_2,z_3,z_4$. Now define:
\be
W_k\equiv \left| \begin{matrix}
f_1 & f_2 & \cdots & f_n\\
 \del_1 f_1 & \del_1 f_2 & \cdots & \del_1f_n\\
\cdots & \cdots  & \cdots & \cdots\\
 \del_1^{k-1} f_1 & \del_1^{k-1} f_2 & \cdots & \del_1^{k-1}f_n\\
 \del_1^{k+1} f_1 & \del_1^{k+1} f_2 & \cdots & \del_1^{k+1}f_n\\
\cdots & \cdots  & \cdots & \cdots\\
\del_1^{n} f_1 & \del_1^{n} f_2 & \cdots & \del_1^{n}f_n
\end{matrix}\right|\qquad \hbox{for}~k=0,1,\cdots n 
\label{Wronskian}
\ee
Note that the row involving $k$th derivatives is omitted in $W_k$. From the fact that the $f_\alpha$ are holomorphic and that crossing symmetry transforms the blocks into linear combinations of themselves, it follows immediately that the Wronskians are holomorphic and crossing-symmetric.
 
Now it is easy to show that the conformal blocks are precisely the independent solutions of a crossing-symmetric differential equation. Supposing $f(z_1)$ is any linear combination of the blocks $f_1,f_2,\cdots, f_n$. Then we have the equation:
\be
\left| \begin{matrix}
f_1 & f_2 & \cdots & f_n &f\\
 \del_1 f_1 & \del_1 f_2 & \cdots & \del_1f_n & \del_1 f\\
\cdots & \cdots  & \cdots & \cdots\\
\del_1^{n-1} f_1 & \del_1^{n-1} f_2 & \cdots & \del_1^{n-1}f_n & \del_1^{n-1} f\\
\del_1^{n} f_1 & \del_1^{n} f_2 & \cdots & \del_1^{n}f_n & \del_1^n f
\end{matrix}\right|=0
\label{eqderiv}
\ee
Expanding by the last column, we have:
\be
\sum_{k=0}^n (-1)^{n-k}\,W_k\,\del_1^k f=0
\ee
This is the desired differential equation. Of course the blocks are unknown, and therefore so are the Wronskians. We only know that they are holomorphic.  But using the operator algebra we can find constraints strong enough to determine them. 

The above equation can be written in monic form:
\be
\del_1^n f_\alpha +\sum_{k=0}^{n-1}\Psi_k\, \del_1^k f_\alpha=0
\label{diffeqn.1}
\ee
where:
\be
\Psi_k=(-1)^{n-k}\frac{W_k}{W_n}
\ee
for all $k$. Although the $W_k$ are holomorphic, $W_n$ in general has zeroes and therefore $\Psi_k$ are meromorphic. One can  easily show that as $z_1\to z_a$ where $a=2,3,4$:
\be
\Psi_k\sim \frac{1}{z_{1a}^{n-k}}\qquad\hbox{where }z_{1a}=z_1-z_a
\label{Psising}
\ee
In principle $\Psi_k$ could also have poles at other points where $W_n$ vanishes, however since the conformal blocks on the plane are non-vanishing away from $z_1\to z_a$, such poles do not exist in this case\footnote{Such ``spurious poles'' can exist for torus correlators \cite{Mathur:1988rx} and also for correlators on the plane with more than two blocks \cite{Fuchs:1992te}.}. The $\Psi_k$ must vanish as $z_1\to \infty$. Indeed, since the number of poles and zeroes on the plane has to be equal, we have:
\be
\Psi_k\to \frac{1}{z_1^{3(n-k)}}\quad \hbox{as~} z_1\to\infty
\ee

To precisely determine the $\Psi_k$, we specialise to the case of small numbers of conformal blocks. Let us consider the 4-point function of a Hermitian primary $\phi_A$ that is known to have two conformal blocks. We assume that the fusion rules of the given CFT allow the conformal families associated to the identity $I$ and one other field $\phi_B$ (the latter may be the same as $\phi_A$ or distinct from it) to flow in the intermediate channel. Because the field $\phi_A$ is Hermitian, one expects that the identity itself necessarily appears in the first channel (otherwise the two-point function of the primary $\phi_A$ with itself would vanish). In the second channel the intermediate state has conformal dimension $h_B$ but this may be either a primary, as we will see in some standard examples, or a secondary at some integer level above a primary, as we will see in the following section.

Let us label the holomorphic conformal dimensions of $\phi_A,\phi_B$ as $h_A,h_B$. The conformal blocks $f_1,f_2$ are associated to the channels $AA\to I\to AA$, $AA\to B\to AA$ respectively. In this case, as $z_1\to z_a$ with $a=2,3,4$, the leading singularity of the conformal blocks is:
\be
f_1\sim \frac{1}{z_{1a}^{2h_A}},\quad f_2\sim \frac{1}{z_{1a}^{2h_A-h_B}}
\ee
and as $z_1\to\infty$ the behaviour is:
\be
f_1\sim \frac{1}{z_{1}^{2h_A}},\quad f_2\sim \frac{1}{z_{1}^{2h_A}}
\ee
From this we deduce that as $z_1\to z_a$ the Wronskian $W_2$ behaves as:
\be
W_2\sim \frac{1}{z_{1a}^{4h_A-h_B+1}}
\ee
while as $z_1\to\infty$, one has:
\be
W_2\sim \frac{1}{z_{1}^{4h_A+2}}
\ee
Requiring that the total singularity be zero, we get:
\be
12h_A-3h_B+3=4h_A+2
\label{Riemannid}
\ee
We call this the Riemann identity. Eliminating $h_B$ using the above equation, the Wronskian $W_2$ is of the form:
\be
W_2=  \kappa(z_2,z_3,z_4)\,  z_{12}^{-\frac23(2h_A+1)}
z_{13}^{-\frac23(2h_A+1)}z_{14}^{-\frac23(2h_A+1)}
\label{wronskiantwo}
\ee
where the pre-factor $\kappa$ is independent of $z_1$.

Let us consider a few simple examples to illustrate this point. The two-character non-unitary ``Lee-Yang'' CFT has a single nontrivial primary $\phi$ with $h=-\frac15$. The fusion rule is $\phi\times \phi = I+\phi$. Thus $h_A=h_B=-\frac15$. It is then easy to verify that \eref{Riemannid} holds. Next consider the spin field $\sigma$ of the Ising model. The relevant fusion rule is $\sigma\times\sigma = I+\epsilon$. Thus $h_A=\frac{1}{16}$ and $h_B=\half$ and \eref{Riemannid} is again satisfied. However, models are known where a primary does not flow in the intermediate channel and the leading singularity comes from a secondary. When this happens, the equality \eref{Riemannid} does not hold and one should replace $h_B$ in it by the dimension of the appropriate secondary. 
For example in \cite{Mathur:1988rx}, 4-point correlators in the SU(N) WZW model were studied. It was argued there that the obvious primary cannot flow for reasons of group theory, and the leading singularity comes from a current-algebra secondary. Then the analogue of the relation \eref{Riemannid} is not satisfied. Later we will see that in the Baby Monster CFT, again it is not the primary that flows in the second channel, but instead  the leading singularity comes from a secondary in the block corresponding to $\phi_B$. However, all these are theories with degenerate (multi-component) primaries, and we will address that class in subsequent sections.

The differential equation for a pair of conformal blocks is:
\be
\del_1^2 f +\Psi_1\del_1 f+\Psi_0f=0
\ee
From the definition of the Wronskians, one can easily verify that $W_1=\del_1 W_2$ (Abel's relation). Hence, from \eref{wronskiantwo}:
\be
\Psi_1=-\frac{W_1}{W_2}=\frac23(2h_A+1)\sum_{i=2,3,4}\frac{1}{z_{1i}}
\ee
This agrees with \eref{Psising}, and we have additionally determined the coefficient. 

Next, we must determine $\Psi_0$. From \eref{Psising} this has a leading double pole in each of the $z_{1i}$. The coefficients can be determined from the known singular behaviour of the individual conformal blocks. One finds:
\be
\Psi_0=-\frac23 h_A(2h_A+1)\sum_{i=2,3,4}\frac{1}{z_{1i}^2}+(\hbox{lower poles in }z_{1i})
\label{psizero}
\ee
The form of the lower pole terms in $\Psi_0$ is determined by the fact that they should depend on differences $z_{ij}$ by translational invariance, and must be non-singular as any two of  $z_2,z_3,z_4$ approach each other. Also on dimensional grounds they must scale like two inverse powers of $z_i$. Finally they must respect total symmetry of the correlator under permutations of $z_2,z_3,z_4$. Hence this term must be of the form:
\be
P\left(\frac{1}{z_{12}z_{13}}+\frac{1}{z_{12}z_{14}}+\frac{1}{z_{13}z_{14}}\right)
\ee
where $P$ is a constant. We now require that when $z_2\to z_4$, the function $z_{13}^{-2h_A}z_{24}^{-2h_A}$ is a solution of the differential equation. As a result we find that:
\be
P=\frac{4h_A}{3}(2h_A+1)
\label{Pvalue}
\ee
Hence the complete expression for $\Psi_0$ is:
\be
\Psi_0=\frac{-2h_A}{3}(2h_A+1)\Bigg(\sum_{i=2,3,4}\frac{1}{z_{1i}^2}-2\sum_{i<j}\frac{1}{z_{1i}z_{1j}}\Bigg)
\ee
and the differential equation becomes:
\be
\begin{split}
&\del_1^2 f + \frac23(2h_A+1)\left(\frac{1}{z_{12}}+\frac{1}{z_{13}}+\frac{1}{z_{14}}\right)\del_1 f\\
&\qquad -\frac{2h_A}{3}(2h_A+1)\left(\frac{1}{z_{12}^2}+ \frac{1}{z_{13}^2}+ \frac{1}{z_{14}^2}\right)f\\
&\qquad +\frac{4h_A}{3}(2h_A+1)\left(\frac{1}{z_{12}z_{13}}+\frac{1}{z_{12}z_{14}}+\frac{1}{z_{13}z_{14}}\right)f=0
\end{split}
\label{twoblockeqn}
\ee

At this point it is worth emphasising an essential difference between the approach described here and that pioneered by \cite{Belavin:1984vu}. In the latter approach, one initially uses null vectors to find a {\em partial} differential equation involving derivatives in all four arguments $z_i$ of the correlator. Conformal invariance is then used to rewrite the desired correlator as some standard factors times a function of the cross-ratio:
\be
z\equiv \frac{z_{12}z_{34}}{z_{14}z_{32}}
\label{crossratio}
\ee
In this way the original PDE can be converted to an ordinary differential equation in $z$. However, in our method we do not need to introduce the cross-ratio at the outset, and we arrive directly at an ordinary differential equation -- \eref{twoblockeqn} -- in one of the variables, say $z_1$, with coefficients that depend on the remaining points. By taking $z_2,z_3,z_4$ to $0,1,\infty$ respectively, we then end up with an ordinary differential equation in $z_1$. This is the stage at which we rename $z_1$ as $z$. The reconstruction of the correlation function, as a function of all four arguments, then proceeds precisely as in \cite{Belavin:1984vu} by treating $z$ as the cross-ratio. 

We will label the solutions of the above differential equation as $f_\alpha(z)$ and refer to them as conformal blocks. Next, we define:
\be
G(z,\zbar)=\sum_{\alpha=1,2}f_\alpha(z)\fbar_\alpha(\zbar)
\ee
Finally to recover the original correlation function $\cG(z_i,\zbar_i)$ from these blocks we use:
\be
\cG(z_i,\zbar_i)=(z_{14}z_{32}\zbar_{14}\zbar_{32})^{-2 h_A}   G(z,\zbar)
\label{calgandg}
\ee
where $z$ has been replaced by its expression in terms of the $z_i$, \eref{crossratio}.

Taking the limit $z_2\to 0, z_3\to 1, z_4\to\infty$ and relabeling $z_1\to z$ we find:
\be
\del_z^2 f + \frac23(2h_A+1)\left(\frac{1}{z}+\frac{1}{z-1}\right)\del_z f
 -\frac23 h_A(2h_A+1)\left(\frac{1}{z}- \frac{1}{z-1}\right)^2f=0
\label{mastereq}
\ee
\eref{mastereq} is a master equation that determines the correlators for {\em all} four-point functions having two conformal blocks, as long as the primaries have no degeneracy. In particular, this equation agrees perfectly with Eq.(5.19) of \cite{Belavin:1984vu} once we take $(z_2,z_3,z_4)\to(0,1,\infty)$ in the latter. Thus we have reproduced the differential equation for two-block 4-point correlators of identical fields in all minimal models! These are the correlation functions involving the ``degenerate'' fields traditionally labelled $\phi_{1,2}$ or $\phi_{2,1}$. However, we have not used the existence or structure of null vectors. As we will see, this enables the method to extend beyond minimal models. 

To solve \eref{mastereq} we make the substitution:
\be
f(z)=\Big(z(1-z)\Big)^{-2h_A} k(z)
\label{fksub}
\ee
and obtain the hypergeometric equation:
\be
z(1-z)\,\del_z^2 k + \frac23(1-4h_A)(1-2z)\,\del_z k + \frac43 h_A(1-4h_A)\,k=0
\ee
Comparing with the standard form of the equation:
\be
z(1-z)\,\del_z^2 k + (c-(a+b+1)z)\,\del_z k -ab\,k=0
\ee
we find that the parameters $a,b,c$ are given by the following relations:
\be
c=\sfrac23(1-4h_A), \qquad a+b+1 = \sfrac43(1-4h_A),\qquad ab= -\frac43 h_A(1-4h_A)
\ee
from which we get:
\be
a=\frac{1-4h_A}{3},\qquad b=-4h_A,\qquad c=\sfrac23(1-4h_A)
\ee
Inserting the well-known solutions we have:
\be
\begin{split}
k_1(z)&={}_2F_1\left(\sfrac13(1-4h_A),-4h_A;\sfrac23(1-4h_A);z\right)\\[2mm]
k_2(z) &= \cN z^{\frac{8h_A+1}{3}} {}_2F_1\left(\sfrac13(1-4h_A),\sfrac23(1+2h_A);\sfrac43(1+2h_A);z\right)
\end{split}
\label{solution}
\ee
where $\cN$ is a normalisation factor. To obtain the actual conformal blocks we must normalise them correctly and restore the power of $z(1-z)$ removed in \eref{fksub}. The first block is already normalised for the following reason. As $z_1\to z_2$, we have $z\to 0$. Now in this limit the original correlator goes like $\frac{1}{|z_{12}z_{34}|^{4h_A}}$ and this behaviour is already taken care of  by the prefactor in Eqs.(\ref{fksub}) and (\ref{calgandg}). Hence we must have $k_1(z)\to 1$ as $z\to 0$, which is indeed the case. Hence it only remains to compute the normalisation of the second block. By performing the crossing transformation $z\to 1-z$ and requiring the normalised blocks to transform by a unitary matrix, which renders the full correlator crossing-invariant, we can determine this normalisation ${\cal N}$. This is carried out in the Appendix.

\subsection{Examples: Ising Model}

Here we specialise to the Ising model. Let us start with the spin field, for which $h_A=\frac{1}{16}$. We know that  $h_B=\half$ (the energy operator) and this is confirmed by solving \eref{Riemannid}. Inserting the value of $h_A$ into \eref{mastereq} one finds the equation:
\be
\del_z^2 f+\frac34 \left(\frac{1}{z}+\frac{1}{z-1}\right)\del_z f -\frac{3}{64}\left(\frac{1}{z}- \frac{1}{z-1}\right)^2 f=0
\label{bpzeqn}
\ee
in perfect agreement with Eq.(E.21) of \cite{Belavin:1984vu}.

Following the procedure in the previous section, we find the conformal blocks to be:
\be
\begin{split}
f_1(z)&=\big(z(1-z)\big)^{-\frac{1}{8}}{}_2F_1(\sfrac{1}{4},-\sfrac14;\sfrac12;z)\\ 
f_2(z)&=\cN \big(z(1-z)\big)^{-\frac{1}{8}} z^{\frac{1}{2}}{}_2F_1(\sfrac{3}{4},\sfrac{1}{4};\sfrac{3}{2};z)
\end{split}
\ee
In this special case the hypergeometric functions reduce to elementary functions:
\be
\begin{split}
{}_2F_1(\sfrac{1}{4},-\sfrac14;\sfrac12;z)&=\shalf\Big((1+\sqrt{z})^\half +(1-\sqrt{z})^\half\Big)\\
z^{\frac{1}{2}}{}_2F_1(\sfrac{3}{4},\sfrac{1}{4};\sfrac{3}{2};z)&=
\Big((1+\sqrt{z})^\half -(1-\sqrt{z})^\half\Big)
\end{split}
\ee
and the desired correlation function of $\sigma$ is therefore:
\be
\langle \prod_{i=1}^4\phi_\sigma(z_i,\zbar_i)\rangle=\frac{1}{|z_{14}z_{32}|^\frac14}\Big( |f_1|^2+\cN^2 |f_2|^2\Big)
\ee
Finally, we use \eref{normnondeg} of Appendix A to find $\cN=\half$.

The other primary of the Ising model is the energy operator $\epsilon$ with dimension $h_A=\half$. We know that its fusion with itself only generates the conformal family of the identity. However, if one inserts $h_A=\half$ in \eref{Riemannid} one finds $h_B=\frac53$ suggesting the presence of a new primary. This is our first example of a ``spurious'' primary, and we will formulate the differential equation as if such a primary exists. The corresponding conformal block will decouple as we show below\footnote{This role of the $h_B=\frac53$ field was also known from the minimal model approach in \cite{Belavin:1984vu}. We recover it in our approach just from the Riemann identity, independently of null-vector considerations or the Kac table.}.  The above information determines the differential equation to be:
\be
\del_z^2 f+\frac43 \left(\frac{1}{z}+\frac{1}{z-1}\right)\del_z f -\frac{2}{3}\left(\frac{1}{z}- \frac{1}{z-1}\right)^2 f=0
\label{halfeqn}
\ee
Following the procedure above we end up with the conformal blocks:
\be
\begin{split}
f_1(z)&=\big(z(1-z)\big)^{-1}{}_2F_1(-2,-\sfrac13;-\sfrac23;z)\\
f_2(z)&={\cal N}\big(z(1-z)\big)^{-1} z^{\frac{5}{3}}{}_2F_1(-\sfrac{1}{3},\sfrac{4}{3};\sfrac{8}{3};z)
\end{split}
\ee
From \eref{normnondeg} the coefficient ${\cal N}$ of the second block  is found to vanish in this case. The first hypergeometric function takes the elementary form:
\be
{}_2F_1(-2,-\sfrac13;-\sfrac23;z)=1-z+z^2
\ee
As a result, the correlation function of the dimension $(\half,\half)$ operator $\epsilon$ is:
\be
\langle \prod_{i=1}^4\phi_\epsilon(z_i,\zbar_i)\rangle=\frac{1}{|z_{14}z_{32}|^2}\frac{1}{|z(1-z)|^2}\Big|1-z+z^2\Big|^2
\ee
This is easily seen to agree with the more manifestly symmetric expression that arises by writing $\epsilon=\psi\psibar$ and computing the correlator by Wick's theorem:
\be
\langle \prod_{i=1}^4\phi_\epsilon(z_i,\zbar_i)\rangle=\left|
\frac{1}{z_{12}z_{34}}-\frac{1}{z_{13}z_{24}}+\frac{1}{z_{14}z_{23}}\right|^2
\ee

\subsection{Examples: Minimal models}

More generally, the field $\phi_{r,s}$ in an arbitrary unitary minimal model with $(p,p')=(m+1,m)$ has dimension:
\be
h_{r,s}=\frac{((m+1)r-ms)^2-1}{4m(m+1)}
\ee
If we choose $(r,s)=(1,2)$ then the fusion rules produce the identity and $\phi_{1,3}$. Thus the four-point correlator for this field always has two conformal blocks.  We easily find that:
\be
h_{1,2}=h_A=\frac{m-2}{4(m+1)},\quad h_{1,3}=h_B=\frac{m-1}{m+1}
\ee
These values are seen to satisfy \eref{Riemannid}. Inserting $h_A$ into \eref{mastereq} we reproduce Eq.(8.71) of \cite{DiFrancesco:1997nk} for arbitrary $m$, where $t$ of that equation is $\frac{m}{m+1}$. As above, the solution can again be expressed in terms of hypergeometric functions and the structure constants fixed by crossing symmetry to obtain the correlation function.

Next consider the field $\phi_{2,1}$ which fuses into the identity and $\phi_{3,1}$. The dimensions this time are:
\be
h_{2,1}=h_A=\frac{m+3}{4m},\quad h_{3,1}=h_B=\frac{m+2}{m}
\ee
Again these values satisfy \eref{Riemannid} and upon inserting $h_A$ into \eref{mastereq} we reproduce Eq.(8.71) of \cite{DiFrancesco:1997nk} where $t$ of that equation is now $\frac{m+1}{m}$. This equation is solved in the same manner as indicated above. 

\section{Wronskian method for RCFT correlators: multi-component primaries}
\label{Wronskian-multiple}

Many CFT's have primary fields with multiple components. For example, in WZW models the primaries transform in representations of the Kac-Moody algebra. However there are also models without any Kac-Moody algebra where the primaries have multiplicities -- one such example is the Baby Monster CFT (\cite{Hoehn:thesis,Hoehn:Baby8,Hampapura:2016mmz}. In this case the multiplicity is ascribed to the transformation property under representations of a discrete group, the Baby Monster group. We will attempt a unified description of all such cases. We will consider two sub-cases, for real and complex primaries respectively. In the latter case each primary must appear along with its complex conjugate in order for the correlation function to be non-zero. On the way we encounter a significant complication relative to the cases addressed in the previous section, which was noted in specific cases in \cite{Mathur:1988rx,Mathur:1988gt} and investigated in a somewhat more general context in \cite{Fuchs:1989pp}: when a primary has a multiplicity, the indices must be combined pairwise into definite representations that flow in the intermediate channel, and this creates ``selection rules'' which influence the singular behaviour of the conformal blocks, the Wronskian and the differential equation. As mentioned previously, these rules can forbid the flow of a primary in specific intermediate channels and the leading singularity comes from a secondary. 

\subsection{Real primaries} 

We first consider the correlator of four identical primary fields $\phi_{a,b}(z,\zbar)$ where $a,b$ run over the left and right degeneracies of the primary respectively\footnote{Here we restrict to left-right symmetric primaries, but the discussion is easily generalised.}. Define:
\be
\cG_{a_1 a_2 a_3 a_4,b_1 b_2 b_3 b_4}(z_i,\zbar_i)=
\langle \phi_{a_1,b_1}(z_1,\zbar_1)\phi_{a_2,b_2}(z_2,\zbar_2)\phi_{a_3,b_3}(z_3,\zbar_3)\phi_{a_4,b_4}(z_4,\zbar_4)\rangle
\ee
Now crossing symmetry does not merely consist of the interchange $z_i\leftrightarrow z_j$, but must be accompanied by an interchange of the indices $a_i,b_i \leftrightarrow a_j,b_j$ as well. 

Now the correlation function is a sum over the modulus-squared of conformal blocks: 
\be
\cG_{a_1 a_2 a_3 a_4,b_1 b_2 b_3 b_4}(z_i,\zbar_i)= \sum_{\alpha=1,2} f_{\alpha,a_1 a_2 a_3a_4}(z_i) \fbar_{\alpha, b_1b_2 b_3 b_4}(\zbar_i)
\label{degencorr}
\ee

On the assumption that the degeneracies labelled by $a_i,b_i$ are due to transformation properties in an irreducible representation $R_A$ of some symmetry group, each of the conformal blocks is a sum over tensor structures:
\be
f_{\alpha,a_1 a_2 a_3a_4}(z_i) =\sum_{R_p\in R_A\otimes R_A}D^{(p)}_{a_1 a_2 a_3 a_4}f_\alpha^{(p)}(z_i)
\label{blocksum}
\ee
with a similar equation for the barred block. Here $R_p$ runs over all irreducible representations contained in $R_A\otimes R_A$, and 
$D_{a_1a_2a_3a_4}^{(p)}$ are tensors that combine $R_A\otimes R_A$ into $R_p$ in the $(a_1a_4)$ and $(a_2a_3)$ channels (and similarly for $\Dbar_{b_1 b_2 b_3 b_4}^{(p)}$)\footnote{This is a different choice from that made in Appendix A of \cite{Mathur:1988rx} and Section 4.2 of \cite{Mathur:1988gt}. The reason will become clear below.}. 

As in the previous section, we take $\phi_A$ to have holomorphic conformal dimension $h_A$. Since this field is real, one of the blocks is associated to the conformal family of the identity operator. The other will be the conformal family of a primary of dimension $h_B$. In special cases this primary can coincide with the original one, and in that case we would have $h_B=h_A$. 

We now seek a set of second-order differential equations for the $f_\alpha^{(p)}$, proceeding as in the previous section but with some notable differences. The key point is that as $z_1\to z_4$, the leading behaviour of the conformal blocks (which was crucial to determine the Wronskian and hence the differential equation) is not necessarily governed by a primary propagating in the intermediate channel. Indeed, due to the fact that the indices $(a_1,a_4)$ and $(a_2,a_3)$ are combined into definite irreducible representations, it can be that group theory forbids the flow of a primary in the intermediate channel, and even of certain secondaries upto a suitably high level. Accordingly, we assume the two blocks $f^{(p)}_\alpha(z_i), \alpha=1,2$ have the following behaviour as $z_1\to z_4$:
\be
f^{(p)}_1(z_i)\sim \frac{1}{z_{14}^{2h_A-n_1}},\quad f^{(p)}_2(z_i)\sim \frac{1}{z_{14}^{2h_A-h_B-n_2}}
\label{asympreal}
\ee
where $n_1,n_2$ are integers $\ge 0$ which label the lowest secondary that can flow in the corresponding channel, given the representation $R_p$. For example if $R_p$ is the identity then we will have $n_1=0$ and if $R_p=R_A$ then we will have $n_2=0$.

To continue, we need the behaviour of the same blocks as $z_1\to z_2,z_3$. Since we have chosen definite representations to flow in the $(14)\to(23)$ channel, we will not have definite representations flowing in the other channels. Therefore in those channels, the primary intermediate state will indeed determine the leading behaviour of the blocks. Thus as $z_1\to z_2$:
\be
f^{(p)}_1(z_i)\sim \frac{1}{z_{12}^{2h_A}},\quad f^{(p)}_2(z_i)\sim \frac{1}{z_{12}^{2h_A-h_B}}
\ee 
The same equation also holds with $z_2$ replaced by $z_3$. Finally, at infinity we have the usual behaviour:
\be
f^{(p)}_1(z_i)\sim \frac{1}{z_1^{2h_A}},\qquad f^{(p)}_2(z_i)\sim \frac{1}{z_1^{2h_A}}
\ee
As a result, the Wronskian $W_2$ behaves as:
\be
\begin{split}
&z_1\to z_2\!:~ W_2\sim \frac{1}{z_{12}^{4h_A-h_B+1}} \\
&z_1\to z_3\!:~ W_2\sim \frac{1}{z_{13}^{4h_A-h_B+1}}\\
&z_1\to z_4\!:~ W_2\sim\frac{1}{z_{14}^{4h_A-h_B-n_1-n_2+1}}\\
&z_1\to \infty\!:~ W_2\sim \frac{1}{z_{1}^{4h_A+2}}
\end{split}
\ee
Requiring that the total singularity be zero, we find the Riemann identity for this case to be:
\be
3h_B+n_1+n_2=
8h_A+1
\label{Riemannid-new}
\ee
Note that this determines only the sum $n_1+n_2$ and not the individual values of $n_1,n_2$. It is precisely this pair of individual values, satisfying $n_1+n_2=8h_A-3h_B+1$, that labels the different allowed irreducible representations $R_p$ in \eref{degencorr}. Note also that $8h_A-3h_B$ must therefore be an integer $\ge -1$ in order for the 4-point function of any CFT with the given fusion rules to be non-vanishing. 

Defining $N=n_1+n_2$ and solving for $h_B$, the Riemann identity \eref{Riemannid-new} can be expressed as:
\be
h_B=\frac{8h_A+1-N}{3}
\label{hbsolv}
\ee
This can be used either to find $h_B$ if $N$ is known from group theory, or to find $N$ if $h_B$ is known via fusion rules. Next let us re-label $n_2=n$ where $0\le n\le N$, so $n_1=N-n$. We can now identify $n$ with the index $p$ in \eref{degencorr}. With this labelling, $R_N$ is the identity representation while $R_0=R_A$. Henceforth we will treat $h_A$ and $N$ as the two independent quantities labelling the CFT, with $h_B$ being determined by \eref{hbsolv}. Then, $n$ labels the component of the correlation function that we are computing and
\eref{blocksum} can be written:
\be
f_{\alpha,a_1 a_2 a_3 a_4}(z_i)=\sum_{n=0}^N D_{a_1 a_2a_3a_4}^{(n)}\, f_\alpha^{(n)}(z_i)
\label{degencorrtwo}
\ee

It should be evident that the non-degenerate case discussed in Section \ref{Wronskian-single} corresponds to $N=0$.

Let us now turn to the calculation of the $N+1$ functions $f_\alpha^{(n)}(z_i)$, each corresponding to a pair of holomorphic conformal blocks satisfying:
\be
\del_1^2 f^{(n)}(z_i) +\Psi_1\del_1 f^{(n)}(z_i)+\Psi_0 f^{(n)}(z_i)=0
\ee
Using the singular behaviour of the Wronskian $W_2$ obtained above and eliminating $h_B$ via \eref{hbsolv}, we find:
\be
W_2=  \kappa(z_2,z_3,z_4)\,  z_{12}^{-\frac23(2h_A+1+\frac{N}{2})} 
z_{13}^{-\frac23(2h_A+1+\frac{N}{2})}z_{14}^{-\frac23(2h_A+1-N)}
\label{wronskiantwon}
\ee
It follows that:
\be
\Psi_1=-\frac{\del_1 W_2}{W_2}=
\frac23(2h_A+1-N)
\frac{1}{z_{14}}+
\frac23\Big(2h_A+1+\frac{N}{2}\Big)
\left(\frac{1}{z_{12}}+\frac{1}{z_{13}}\right)
\label{Psione}
\ee
Finally we determine $\Psi_0$. This takes the form: 
\be
\Psi_0=
-\frac13 (2h_A-N+n)(2h_A+1-N-3n)
\frac{1}{z_{14}^2}
-\frac23 h_A(2h_A+1-N)
\left(\frac{1}{z_{12}^2}+ \frac{1}{z_{13}^2}\right) + {\tPsi_0}(z_{1i})
\label{Psizero.1}
\ee
where $\tPsi_0$ has only simple poles in the $z_{1i}$. Using dimensional analysis and the residual symmetry $z_2\leftrightarrow z_3$ we can write:
\be
\tPsi_0= Q\left(\frac{1}{z_{12}z_{14}}+\frac{1}{z_{13}z_{14}}\right)+R\frac{1}{z_{12}z_{13}}
\label{Psizero.2}
\ee
where $Q$ and $R$ are constants. We determine them by taking suitable limits and comparing double poles. For $Q$, we consider the limit $z_1\to z_3, z_2\to z_4$. In this limit, the differential equation is solved by $z_{13}^{-2h_A}z_{24}^{-2h_A}$. Similarly we take the limit $z_1\to z_4,z_2\to z_3$ which determines $R$. Hence we find:
\be
\begin{split}
Q &=\frac{4h_A}{3} (2 h_A +  1 + 2 n - \sfrac32 N) - \frac13(N - n) (1 + 3 n - N)\\[2mm]
R&=\frac{4h_A}{3}(2h_A+1-2n+N)+\frac13(N-n)(1+3n-N)
\label{qrval}
\end{split}
\ee
One can verify that for $n=N=0$, these expressions reduce to $Q=R=P$ where $P$ was evaluated in \eref{Pvalue} above.

Now we combine Eqs.(\ref{Psione}-\ref{qrval}). As in the previous section, we take the limits $z_2\to 0,z_3\to 1, z_4\to \infty$\footnote{This too is different from the limit taken in \cite{Mathur:1988gt}. In that reference, the limits were chosen to simplify the equation, but in the convention where the group theory factors created a definite representation in the (12) and (34) channels.  Here, however, we have chosen the group theory factors to create a definite representation in the (14) and (23) channels so that we can retain the standard limits for $z_2,z_3,z_4$.}. Denoting $z_1$ by $z$, we finally get the differential equation:
\be
\begin{split}
&\del_z^2 f^{(n)} +\frac23\Big(2h_A+1+\frac{N}{2}\Big)
\left(\frac{1}{z}+\frac{1}{z-1}\right)\del_z f^{(n)} +
\Bigg\{
-\frac23 h_A(2h_A+1-N)
\left(\frac{1}{z^2}+ \frac{1}{(z-1)^2}\right)\\
& \qquad+\Big(\frac{4h_A}{3}(2h_A+1-2n+N)+\frac13(N-n)(1+3n-N)\Big)\frac{1}{z(z-1)}
\Bigg\}
f^{(n)}=0
\end{split}
\label{mastereq.2}
\ee
Thus we have arrived at a set of $N+1$ master equations (one for  each value of $n$ lying between 0 and $N$) for the four-point functions of identical, real fields in {\em any} RCFT as long as the correlator has only two conformal blocks.

To solve each equation, we make the substitution as before:
\be
f(z)=\Big(z(1-z)\Big)^{-2h_A}k(z)
\label{solvsub}
\ee
The power in this equation does not depend on $n$. This happens because the coefficients of the pure single and double poles in the above equation are both independent of $n$ (only the coefficient of the mixed pole depends on $n$). The result is:
\be
z(1-z)\del_z^2 k +\Big(\frac23(1-4h_A)+\frac{N}{3}\Big)(1-2z)\del_z k
-\frac13(4h_A-N+n)(4h_A-1+N-3n)k=0
\ee
Comparing with the standard form of the hypergeometric equation, we get:
\be
\begin{split}
a &=\frac13(1-4h_A-N+3n)\\
b &=-4h_A+N-n\\
c &=\frac23(1-4h_A)+\frac{N}{3}
\end{split}
\ee
As a result, the two conformal blocks for each $n$ are:
\be
\begin{split}
f^{(n)}_1(z)&=\big(z(1-z)\big)^{-2h_A}{}_2F_1\Big(\sfrac13(1-4h_A-N+3n),-4h_A+N-n;\sfrac23(1-4h_A)+\sfrac{N}{3};z\Big)\\[2mm]
f^{(n)}_2(z) &= \cN^{(n)}\big(z(1-z)\big)^{-2h_A}z^{\frac{8h_A+1-N}{3}}\\
&\qquad\qquad \times {}_2F_1\left(\sfrac23(1+2h_A-N)+n,\sfrac{1-4h_A+2N}{3}-n;\sfrac43(1+2h_A)-\sfrac{N}{3};z\right)
\end{split}
\label{degsolution}
\ee
The monodromy matrix and normalisation constants $\cN_n$ for this case are computed in Appendix B.

\subsection{Complex primaries}

When $\phi_A$ is complex, we label it $\phi_{a,b}(z,\zbar)$ and its complex conjugate $\phi_{\abar,\bbar}(z,\zbar)$. 
The correlation function of interest now has two complex fields and two complex conjugates:
\be
G_{a_1 \abar_2 \abar_3 a_4,b_1\bbar_2 \bbar_3 b_4}(z_i,\zbar_i)=
\langle \phi_{a_1,b_1}(z_1,\zbar_1)\phibar_{\abar_2,\bbar_2}(z_2,\zbar_2)\phibar_{\abar_3,\bbar_3}(z_3,\zbar_3)\phi_{a_4,b_4}(z_4,\zbar_4)\rangle
\ee

The analogues of Eqs.(\ref{degencorr}) and (\ref{blocksum}) are now:
\be
\cG_{a_1 \abar_2 \abar_3 a_4,b_1 \bbar_2 \bbar_3 b_4}(z_i,\zbar_i)= \sum_{\alpha=1,2} f_{\alpha,a_1 \abar_2 \abar_3a_4}(z_i) \fbar_{\alpha, b_1 \bbar_2 \bbar_3 b_4}(\zbar_i)
\label{degencorrcomp}
\ee
and:
\be
f_{\alpha,a_1 \abar_2 \abar_3 a_4}(z_i) =\sum_{R_p\in R_A\otimes R_A}D^{(p)}_{a_1 \abar_2 \abar_3 a_4}f_\alpha^{(p)}(z_i)
\label{blocksumcomp}
\ee

The reason to arrange the fields in the order $\phi\phibar\phibar\phi$ is related to our choice in the previous section that the group theory factors give definite representations in the (14) channel. When we do this in the complex case and let the last field be $\phi$, the answer is symmetric under interchange of the remaining two $\phibar$ fields. Hence the resulting differential equation will be symmetric under $z_2\leftrightarrow z_3$ which is a useful simplification.

We now have two different fusion rules:
\be
\phi_A\otimes \phibar_A=\II+\phi_B,\qquad \phi_A\otimes \phi_A=\phi_C+\phi_D
\ee
and this will significantly influence the final result, though the method remains the same as in the previous sections.

Again we assume that $D$ and $\Dbar$ are chosen to combine the representations of the holomorphic/anti-holomorphic part of the fields 
$\phi_{a_1,b_1}$ and $\phi_{a_4,b_4}$ into definite representations of the algebra. The representations of 
$\phibar_{\abar_2,\bbar_2}$ and $\phibar_{\abar_3,\bbar_3}$ will also combine into that same representation. This only restricts what can flow in the intermediate channel when we fuse $z_1\to z_4$, while generic representations will flow in the other channels. It follows that in the various coincident limits, the blocks behave as follows:
\begin{equation}
\begin{split}
z_1\to z_2\!: \qquad &f_1^{(n)}\sim z_{12}^{-2h_A}, \qquad\qquad\qquad f_2^{(n)}\sim z_{12}^{-2h_A+h_B} \\ 
z_1\to z_3\!: \qquad &f_1^{(n)}\sim z_{13}^{-2h_A}, \qquad\qquad\qquad f_2^{(n)}\sim z_{13}^{-2h_A+h_B} \\
z_1\to z_4\!: \qquad &f_1^{(n)}\sim z_{14}^{-2h_A+h_C+N-n} \qquad  f_2^{(n)}\sim z_{14}^{-2h_A+h_D+n} 
\end{split}
\end{equation}
Incorporating the standard behaviour at infinity, the Riemann identity is now:
\be
2h_B+h_C+h_D=8h_A+1-N
\label{Riemannid-complex}
\ee
As an example, for the fundamental primary of SU(M)$_k$ we have\footnote{We label the SU algebras by the integer $M$ to avoid confusion with $N$ and $n$ which were introduced around \eref{hbsolv}.}:
\be
\begin{split}
&h_A=\frac{M^2-1}{2M(M+k)},\qquad\qquad h_B=\frac{M}{M+k}\\
&h_C=\frac{(M-2)(M+1)}{M(M+k)},\qquad h_D=\frac{(M+2)(M-1)}{M(M+k)}
\end{split}
\ee
In this case it is easily seen that $N=1$ always. From group theory, this corresponds to the fact that $\phi_C$ (resp. $\phi_D$) are the symmetric (resp. antisymmetric) representations contained in the tensor product of the fundamental with itself, but when we combine the group indices into symmetric/antisymmetric combinations, one or the other of these primaries cannot flow and has to be replaced by the secondary immediately above it. Notice also that \eref{Riemannid-complex} above corrects a typo in Eq.(A.5) of \cite{Mathur:1988rx}, whose RHS should be $-4h_g-2$ in the notation of that paper.

Following the same procedure as in the previous section, we are now led to the differential equation:
\begin{equation}
\begin{split}
&\partial_{z_{1}}^2 f^{(n)}  + \Bigg[ (4h_A-h_B+1)\Big(\frac{1}{z_{12}}+\frac{1}{z_{13}}\Big)
+(4h_A-h_C-h_D+1)\frac{1}{z_{14}}
\Bigg]\partial_{z_1}f^{(n)} \\
&+ \Bigg[ -2h_A(-2h_A+h_B)\Big(\frac{1}{z_{12}^2}+\frac{1}{z_{13}^2}\Big)
+(-2h_A+h_C+N-n)(-2h_A+h_D+n)\frac{1}{z_{14}^2}\\
&+\Big((-2h_A+h_C+N-n)(-2h_A+h_D+n)-2h_A(-2h_A+h_B)\Big)\Bigg(\frac{1}{z_{12}z_{14}}+\frac{1}{z_{13}z_{14}}\Bigg)\\
&+\bigg(2h_A(10h_A-2h_B-h_C-h_D-N+2)\\
&\qquad\qquad\qquad +(-2h_A+h_C+N-n)(-2h_A+h_D+n)\bigg)\frac{1}{z_{12}z_{13}}\Bigg]
f^{(n)} = 0 
\end{split}
\end{equation}
If we take $N=1, n=0$ and exchange $z_2\leftrightarrow z_4$, this agrees with Eq(A.10) of \cite{Mathur:1988rx} which was derived for precisely this case. However our result above has more general applicability as we will soon see.

At this stage we take the limit $z_2 \rightarrow 0$, $z_3\rightarrow 1$, $z_4\rightarrow \infty$ to recover an ordinary differential equation:
\begin{equation}
\begin{split}
&\partial_z^2 f^{(n)}+ (4h_A-h_B+1)\Big(\frac{1}{z}+\frac{1}{z-1}\Big)\partial_z f^{(n)} +
\Bigg[-2h_A(-2h_A+h_B)\Big(\frac{1}{z^2}+\frac{1}{(z-1)^2}\Big)\\
&+\bigg(2h_A(10h_A-2h_B-h_C-h_D-N+2)+\\
&\qquad \qquad (-2h_A+h_C+N-n)(-2h_A+h_D+n)\bigg)\frac{1}{z(z-1)}\Bigg] f^{(n)} = 0
\end{split}
\label{complexdiff}
\end{equation}   
This is converted into standard hypergeometric form by the substitution \eref{solvsub} and we end up with: 
\begin{equation}
z(1-z)\,\partial_z^2 k^{(n)}+(1-h_B)(1-2z)\,\partial_z k^{(n)}-(4h_A-h_D-n)(4h_A-h_C-N+n)\, k^{(n)}= 0
\end{equation}
Comparing with the standard form, we identify the labels $a,b,c$ of the hypergeometric function to be:
\begin{equation}
\begin{split}
a &= -4h_A+h_D+n \\
b&= -4h_A+h_C+N-n \\
 c= 1-h_B & = \sfrac{1}{2}(1-8h_A+h_C+h_D+N)
\end{split}
\end{equation} 
The solutions are:
\begin{equation}
\begin{split}
f_1^{(n)} &= (z(1-z))^{-2h_A} {}_2F_1 \big(-4h_A+h_D+n, -4h_A+h_C+N-n ; 1-h_B;z\big)  \\
f_2^{(n)} &= \cN^{(n)}(z(1-z))^{-2h_A}z^{h_B}\\
&\qquad\qquad\times {}_2F_1\big(-4h_A+h_B+h_D+n, -4h_A+h_B+h_C+N-n ; 1+h_B; z\big)
\end{split}
\label{complexblocks}
\end{equation}
As a check, this reduces to the real case of the previous subsection if we put $h_C=0$ and $h_D=h_B$.
Finally, the monodromy matrix and normalisation constant are computed in the Appendix C.

\subsection{The Deligne-Cvitanovi\'c Exceptional Series}

In \cite{Mathur:1988na}, a class of 2d Rational CFT's was completely classified: those with precisely two characters and no zeroes for the leading Wronskian. The result was 7 WZW models, having central charges satisfying $0<c<8$, and one non-unitary minimal model -- the so-called Lee-Yang theory. Several years later it was observed by Deligne  that the Lie algebras for precisely these CFT's have special representation-theoretic properties\cite{Deligne:1,Deligne:2}.  Following current literature (e.g.\cite{Beem:2017ooy}) we refer to this as the Deligne-Cvitanovi\'c exceptional series\footnote{As an aside, \cite{Mathur:1988na} also discovered a theory lying ``between'' E$_7$ and E$_8$, which satisfies most axioms of CFT (except that the ``identity'' is degenerate). This matches the so-called E$_{7.5}$ algebra \cite{Landsberg:2004}. We are grateful to Yuji Tachikawa for pointing this out.}. Each of these models has either (i) a single real primary field other than the identity, or (ii) a single complex primary field, (iii) three primaries of the same conformal dimension (due to triality). Let us  apply the above method to each of them in turn.

While the bootstrap method as implemented in this paper does not require a priori knowledge of the current algebra and spectrum of integrable primaries for WZW models, we do have that knowledge for all elements of the Deligne series. This fact will be useful for us to verify our results, and in a few difficult cases will have to be used to supplement the bootstrap information. Some formulae relevant for WZW models are as follows. The Virasoro central charge is determined by:
\be
c=\frac {k\,{\rm dim}\cG}{k+g}
\label{centr}
\ee
where $\cG$ is the algebra, $k$ is the level and $g$ the dual Coxeter number. The conformal dimension of a primary in the irreducible representation $\cR$ is:
\be
h=\frac{\vlambda\cdot(\vlambda+2\vrho)}{2(k+g)}
\label{confr}
\ee
where $\vlambda,\vrho$ are respectively the weight of the representation and the Weyl vector. There is a convenient way to rewrite this. We have:
\be
\vlambda\cdot(\vlambda+2\rho)=\frac{{\rm dim}\cG}{{\rm dim}\cR}\, \ell(\cR)
\label{indr}
\ee
where the index $\ell(\cR)$ of the representation is relatively easy to look up, for example in \cite{feger:2012}. Inserting Eqs.(\ref{centr}, \ref{indr}) into \eref{confr} we find:
\be
h=\frac{c\,\ell(\cR)}{2\,{\rm dim}(\cR)}
\label{confdim}
\ee
For example in SU(2) we have:
\be
\ell(j)=\sfrac23 j(j+1)(2j+1)
\ee
where $j$ is the spin, from which one recovers the well-known formula:
\be
h=\frac{j(j+1)}{k+2}
\ee

\subsubsection{SU(2)$_1$}

This theory has a single primary field of dimension $h_A=\sfrac14$ in the doublet representation of SU(2) and fusion rules $\phi_A\times \phi_A=\II$. Thus there must be a ``spurious'' primary, as was the case for the Ising model considered earlier. From \eref{hbsolv} we see that $h_B=1-\frac{N}{3}$. Now since a spin-$\half$ primary fuses with itself into two representations, of spin 0 and spin 1 respectively, we have the possibility of $n=0$ and $n=1$. It follows that $N=1$ and the spurious primary has dimension $h_B=\sfrac23$. Inserting $h_A=\frac14$ and $N=1$ in \eref{gensol} and \eref{normdegreal} gives:
\be
\begin{split}
f^{(n)}_1(z)&=\big(z(1-z)\big)^{-\frac12}{}_2F_1\Big(n-\sfrac13, -n;\sfrac13;z\Big)\\[2mm]
f^{(n)}_2(z) &= \cN^{(n)}z^\frac23\big(z(1-z)\big)^{-\frac12}\, {}_2F_1\left(n+\sfrac13,\sfrac23-n;\sfrac53;z\right)\\[2mm]
\cN^{(n)}&=\frac{\Gamma \left(-\frac{2}{3}\right)}{\Gamma \left(\frac{2}{3}\right)}
   \sqrt{\frac{\Gamma \left(\frac{4}{3}-n\right) \Gamma
   (n+1)}{\Gamma \left(n-\frac{1}{3}\right) \Gamma
   (-n)}}
\end{split}
\ee
We see that for the allowed values $n=0,1$, the normalisation factor vanishes. This means that the second block decouples and we have only a single conformal block for each $n$, namely $f^{(n)}_1(z)$, as expected from the fusion rules. Note that if we had instead assumed $N=0$ or $N=2$, this decoupling would not have taken place, confirming that we must have $N=1$. The surviving blocks can be written in terms of elementary functions as:
\be
\begin{split}
f^{(0)}_1(z)&=(z(1-z))^{-\frac{1}{2}}=(z(1-z))^{\frac{1}{2}}\Big(\frac{1}{z}+\frac{1}{1-z}\Big)\\
f^{(1)}_1(z)&=(z(1-z))^{-\frac{1}{2}}(1-2z)=(z(1-z))^{\frac{1}{2}}\Big(\frac{1}{z}-\frac{1}{1-z}\Big)
\end{split}
\ee

The tensor structures corresponding to $n=0,1$ are simply the symmetric/anti-symmetric combinations:
\be
\begin{split}
D^{(0)}_{a_1a_2a_3a_4}&=\shalf(\delta_{a_1a_2}\delta_{a_3a_4}+\delta_{a_1a_3}\delta_{a_2a_4})\\
D^{(1)}_{a_1a_2a_3a_4}&=\shalf(\delta_{a_1a_2}\delta_{a_3a_4}-\delta_{a_1a_3}\delta_{a_2a_4})\\
\Dbar^{(0)}_{b_1b_2b_3b_4}&=\shalf(\delta_{b_1b_2}\delta_{b_3b_4}+\delta_{b_1b_3}\delta_{b_2b_4})\\
\Dbar^{(1)}_{b_1b_2b_3b_4}&=\shalf(\delta_{b_1b_3}\delta_{b_2b_4}-\delta_{b_1b_2}\delta_{b_3b_4})
\end{split}
\ee
In terms of these, the final four-point function takes the form:
\be
\begin{split}
G(z,\bar{z})&=(z\bar{z}(1-z)(1-\bar{z}))^{\frac{1}{2}}\Bigg(D^{(0)}\left(\frac{1}{z}+\frac{1}{1-z}\right)
+ D^{(1)}\left(\frac{1}{z}-\frac{1}{1-z}\right)\Bigg)\times\\
&\qquad\qquad\Bigg(\Dbar^{(0)}\left(\frac{1}{\zbar}+\frac{1}{1-\zbar}\right)
+ \Dbar^{(1)}\left(\frac{1}{\zbar}-\frac{1}{1-\zbar}\right)\Bigg)\\
&= (z\bar{z}(1-z)(1-\bar{z}))^{\frac{1}{2}}\Big(\frac{I_1}{z}+\frac{I_2}{1-z}\Big)\Big(\frac{{\bar I}_1}{\zbar}+\frac{{\bar I}_2}{1-\zbar}\Big)
\end{split}
\ee
where $I_1=D^{(0)}+D^{(1)}, I_2=D^{(0)}-D^{(1)}$.

As a check of our result, we can compare the above blocks with those computed in Eqs.\,(4.10a,\,4.10b) of \cite{Knizhnik:1984nr}, evaluated for SU(2) 
and there is perfect agreement.

\subsubsection{SU(3)$_1$}

There are two primaries, the 3 and ${\bar 3}$. In this case we must consider the correlator for two $3$'s and two ${\bar 3}$'s. We have $h_A=\frac13$. The fusion rules, as discussed above, are:
\be
\phi_A\otimes\phibar_A =\II+\phi_B,\qquad \phi_A\otimes\phi_A =\phi_C+\phi_D
\ee
with $h_B=\frac34, h_C=\frac13$ and $h_D=\frac56$. From the study of WZW models we know that the primaries corresponding to $h_B$ and $h_D$ are absent from the theory, these are respectively the 8 and the 6 of SU(3) and are decoupled at level $k=1$. Hence we should find just one conformal block at the end. With our method, this fact emerges on its own from computation of the normalisation, just as we saw in the previous case. Indeed, after verifying that the Riemann identity \eref{Riemannid-complex} is satisfied for $N=1$, we insert the conformal dimensions into \eref{complexblocks} and find:
\be
\begin{split}
f_1^{(n)} &= (z(1-z))^{-\frac23} {}_2F_1 \big(n-\shalf, -n ; \sfrac14;z\big)  \\
f_2^{(n)} &= \cN^{(n)}z^\frac34 (z(1-z))^{-\frac23} {}_2F_1\big(n+\sfrac14, \sfrac34-n ;\sfrac74; z\big)
\end{split}
\ee
From \eref{normdegreal} the normalisation factor is found to be:
\be
\cN^{(n)}=\left|\frac{\Gamma\left(-\frac{3}{4}\right)}{\Gamma\left(\frac{3}{4}\right)} \sqrt{\frac{\Gamma\left(\sfrac32-n\right)\Gamma(1+n)}{\Gamma\left(n-
\shalf\right)\Gamma(-n)}}\right|
\ee
As expected, this vanishes for both $n=0$ and $n=1$, confirming that there is a single conformal block for SU(3), namely $f_1^{(n)}(z)$.

The surviving blocks for the four-point function can again be expressed in terms of elementary functions:
\be
\begin{split}
f^{(0)}_1(z)&= \big(z(1-z)\big)^{-\frac23}=\big(z(1-z)\big)^{\frac{1}{3}}\Big(\frac{1}{z}+\frac{1}{1-z}\Big)\\
f^{(1)}_1(z)&= \big(z(1-z)\big)^{-\frac23}(1-2z)=
\big(z(1-z)\big)^{\frac{1}{3}}\Big(\frac{1}{z}-\frac{1}{1-z}\Big)
\end{split}
\ee

We note in passing that the computation is analogous for all SU(M)$_1$  and the blocks for general M are simply:
\be
\begin{split}
f^{(0)}_1(z)&=\big(z(1-z)\big)^{\frac{1}{\rm M}}\Big(\frac{1}{z}+\frac{1}{1-z}\Big)\\
f^{(1)}_1(z)&=\big(z(1-z)\big)^{\frac{1}{\rm M}}\Big(\frac{1}{z}-\frac{1}{1-z}\Big)
\end{split}
\ee

In every fusion, from the group theory point of view there are two output representations:
\be
\begin{split}
\mathbf{M}\otimes \overline{\mathbf{M}}&=\mathbf{1}\oplus \hbox{\bf Adjoint}\\
\mathbf{M}\otimes {\mathbf{M}}&=\hbox{\bf Sym}\oplus \hbox{\bf Antisym}\\
\end{split}
\ee
which for SU(3) translates into the well-known:
\be
\begin{split}
\mathbf{3}\otimes \overline{\mathbf{3}}&=\mathbf{1}\oplus \mathbf{8}\\
\mathbf{3}\otimes {\mathbf{3}}&=\mathbf{6}\oplus \overline{\mathbf{3}}
\end{split}
\ee
 It is easy to see that the integer $N$ should be one less than the number of output representations, because this gives $N+1$ allowed values of the integer $n$ (which ranges from 0 to $N$) and each of these corresponds to the case where one of the output representations can flow in the intermediate channel. This explains why, as noted in the previous section, $N=1$ for all SU(M)$_1$ theories. It also gives a way of computing $N$ in all WZW models. 

\subsubsection{(G$_2$)$_1$}

The single primary $\phi_A$ is real and has dimension $h_A=\sfrac25$. The fusion rules are $\phi_A\times\phi_A=\II+\phi_A$. \eref{hbsolv} gives $h_B=\sfrac75-\sfrac{N}{3}$. Since the fusion rules imply $h_A=h_B$, we find from \eref{hbsolv} that $N=3$. This is confirmed by the group-theoretic relation that the fundamental of G$_2$ satisfies:
\be
\mathbf{7}\otimes \mathbf{7}=\mathbf{1}\oplus \mathbf{7}\oplus \mathbf{14}\oplus \mathbf{27}
\ee
As discussed above, these four possibilities correspond to $N=3$.

Inserting $h_A=\sfrac25$ and $N=3$ in \eref{gensol} and \eref{normdegreal} gives the conformal blocks and normalisation factors respectively to be:
\be
\begin{split}
f^{(n)}_1(z)&=\big(z(1-z)\big)^{-\frac45}{}_2F_1\Big(n-\sfrac65,\sfrac75-n;\sfrac35;z\Big)\\[2mm]
f^{(n)}_2(z) &={\cal N}^{(n)} \big(z(1-z)\big)^{-\frac45}z^{\frac25}\, {}_2F_1\left(n-\sfrac45,\sfrac95-n;\sfrac75;z\right)\\[2mm]
\cN^{(n)}&=\frac{\Gamma \left(-\frac{2}{5}\right)}{\Gamma
   \left(\frac{2}{5}\right)}
   \sqrt{\frac{\Gamma \left(\frac{11}{5}-n\right) \Gamma
   \left(n-\frac{2}{5}\right)}{\Gamma
   \left(\frac{7}{5}-n\right) \Gamma
   \left(n-\frac{6}{5}\right)}}
\end{split}
\ee

\subsubsection{SO(8)$_1$}

Here there are three different primary fields, all of dimension $h_A=\half$. If we consider correlators of four identical primaries, the fusion rules permit only the identity family to flow -- thus we should have a single conformal block and the other one should be spurious. \eref{hbsolv} gives:
\be
h_B=\frac{5-N}{3}
\ee
Notice that $N=0$ would give $h_B=\frac53$, which gives back the case of the dimension-$\half$ primary of the Ising model that we encountered above. As we saw, one of the blocks then decouples. However in the present case each of the primaries is 8-fold degenerate so $N$ typically has a nontrivial value. Indeed, any of the 8's of SO(8) satisfies $8\otimes 8=1\oplus 28\oplus 35$. Following the previous rule (that $N$ is one less than the number of representations in $\phi_A\otimes\phi_A$) we should have $N= 2$. The corresponding tensor structures are:
\be
\begin{split}
D^{(0)}& = \shalf(\delta_{a_1a_2}\delta_{a_3a_4}+\delta_{a_1a_3}\delta_{a_2a_4})\\
D^{(1)}& = \shalf(\delta_{a_1a_2}\delta_{a_3a_4}-\delta_{a_1a_3}\delta_{a_2a_4})\\
D^{(2)}& = \delta_{a_1a_4}\delta_{a_2a_3}
\end{split}
\label{so8tensor}
\ee

With these values we find that the parameters $(a,b,c)$ of the hypergeometric equation turn out to be $(n-1,-n,0)$ respectively. It is well-known that when $c\le 0$, the solutions of the equation are different from the usual ones. In fact it is simpler in this case to solve the equation directly in terms of elementary functions. We have:
\be
\del_z^2 f^{(n)} +2\left(\frac{1}{z}+\frac{1}{z-1}\right)\del_z f^{(n)} +(n-2)(n+1)\frac{1}{z(z-1)}f^{(n)}=0
\ee
The differential equation is the same for $n=0,1$, and the independent solutions in both cases are simply: $\frac{1}{z}\pm\frac{1}{1-z}$.
Meanwhile for $n=2$ one has $f_1(z)=1$, a constant solution, along with a logarithmic solution. 

As seen from these results, this example is quite special (it appears related to the fact that SO(8)$_1$ has reducible monodromy \cite{Franc:2019}). To find the full correlation function, we plug in the values of $h_A,N$ into \eref{Mmatrixreal} to find the monodromy matrices:
\be
{\cal M}^{(0)}={\cal M}^{(1)}=
\begin{pmatrix}
1 & \frac{1}{\cN}\\
\cN & 1
\end{pmatrix},\qquad 
{\cal M}^{(2)}=
\begin{pmatrix}
-1 & -\frac{2}{\cal N}\Gamma(0)\\
\half\frac{\cN}{\Gamma(0)} & 1
\end{pmatrix}
\ee
Both these matrices are singular. This tells us that we should change basis to diagonalise the monodromy and then keep only the first solution. Among other things, this change of basis leads to the first two tensor structures in \eref{so8tensor}. Comparison with \eref{asympreal} tells us that we must choose:
\be
\begin{split}
f_1^{(0)} & = \left(\frac{1}{z}+\frac{1}{1-z}\right)\\
f_1^{(1)} & = \left(\frac{1}{z}-\frac{1}{1-z}\right)\\
f_1^{(2)} &= 1
\end{split}
\ee
while in each case dropping the other solution. Using:
\be
D^{(0)}f^{(0)}+D^{(1)}f^{(1)}=\frac{1}{z}\delta_{a_1a_2}\delta_{a_3a_4}+\frac{1}{1-z}\delta_{a_1a_3}\delta_{a_2a_4}
\ee
we finally get the correlation function:
\be
G_{a_1a_2a_3a_4,\abar_1\abar_2\abar_3\abar_4}(z,\zbar)=\left| \frac{1}{z}\delta_{a_1a_2}\delta_{a_3a_4}+\frac{1}{1-z}\delta_{a_1a_3}\delta_{a_2a_4}-\delta_{a_1a_4}\delta_{a_2a_3}\right|^2
\ee
which precisely agrees with the answer obtained using the free-fermion description of this theory (as a system of 8 free Majorana fermions with correlated spin structures). 

\subsubsection{(F$_4$)$_1$}

This has the same fusion rules as the (G$_2$)$_1$ theory, and was studied in \cite{Mathur:1988gt}. The primary has $h_A=\frac35$ and corresponds to the fundamental representation of dimension 26. The analysis of \cite{Mathur:1988gt} explicitly identifies $N=4$. This result follows immediately from the Riemann identity \eref{Riemannid-new}, and independently from our rule that $N$ should be one less than the number of irreps produced in:
\be
\mathbf{26}\otimes \mathbf{26}=\mathbf{1}\oplus \mathbf{26}\oplus \mathbf{52}\oplus \mathbf{273}\oplus \mathbf{324}
\ee
From \eref{gensol} and \eref{normdegreal}, the blocks and normalisation are found to be:
\be
\begin{split}
f^{(n)}_1(z)&=\big(z(1-z)\big)^{-\frac65}{}_2F_1\Big(n-\sfrac95,\sfrac85-n;\sfrac25;z\Big)\\[2mm]
f^{(n)}_2(z) &= {\cal N}^{(n)}\big(z(1-z)\big)^{-\frac65}z^{\frac35}\, {}_2F_1\left(n-\sfrac65,\sfrac{11}{5}-n;\sfrac85;z\right)\\[2mm]
\cN_n&=\frac{\Gamma \left(-\frac{3}{5}\right)}{\Gamma
   \left(\frac{3}{5}\right)}
   \sqrt{\frac{\Gamma \left(\frac{14}{5}-n\right) \Gamma
   \left(n-\frac{3}{5}\right)}{\Gamma
   \left(\frac{8}{5}-n\right) \Gamma
   \left(n-\frac{9}{5}\right)}}
\end{split}
\ee
This is a very quick re-derivation of the result of \cite{Mathur:1988gt}, showing the power of our general formula.

\subsubsection{(E$_6$)$_1$}

This theory has a complex primary in the $\mathbf{27}$ of E$_6$ having conformal dimension $\frac23$. The fusion rules are the same as for the SU(3)$_1$ theory. The relevant tensor products are:
\be
\begin{split}
\mathbf{27}\otimes \overline{\mathbf{27}}&=\mathbf{1}\oplus \mathbf{78}\oplus \mathbf{650}\\
\mathbf{27}\otimes \mathbf{27}&=\overline{\mathbf{27}}\oplus \overline{\mathbf{351}}\oplus \overline{\mathbf{351'}}
\end{split}
\ee
From previous discussions, the fact that there are three output primaries in each channel indicates that $N=2$. 

Out of the representations appearing above, note that only the $\mathbf{27}$ and $\overline{\mathbf{27}}$ are genuine primaries of this theory, while the remaining primaries are spurious at $k=1$. Their conformal dimensions, according to \eref{confdim}, are as follows:
\be
\mathbf{78}:~h=\sfrac{12}{13},\quad
\mathbf{351}:~h=\sfrac{50}{39},\quad
\mathbf{351'}:~h=\sfrac{56}{39},\quad
\mathbf{650}:~h=\sfrac{18}{13}
\ee
We now have a puzzle in trying to satisfy the Riemann identity:
\be
2h_B+h_C+h_D+N=8h_A+1
\ee
Here, we have $h_A=\sfrac23$ and $h_C=\frac23$. However, the role of $\phi_B$ is played by two fields, the $\mathbf{78}$ and $\mathbf{650}$, of dimensions $\sfrac{12}{13}$ and $\sfrac{18}{13}$ respectively. Similarly the role of $\phi_D$ is played by two fields, the $\mathbf{351}$ and $\mathbf{351'}$, of dimensions $\sfrac{50}{39}$ and $\sfrac{56}{39}$ respectively. In principle we should therefore extend our approach from two to three conformal blocks in each channel, which entails re-doing the entire differential equation approach for this case. However there is an intriguing numerological fact that reproduces the result we already guessed for $N$, namely $N=2$. 
Suppose we set $h_B$ to be the average dimension of the $\mathbf{78}$ and $\mathbf{650}$, which is $\sfrac{15}{13}$. Similarly we set $h_D$ to be the average of the $\mathbf{351}$ and $\mathbf{351'}$
which is $\sfrac{53}{39}$. Now insert these values of $h_A,h_B,h_C,h_D$ into the Riemann identity to get:
\be
N=1+8\times\sfrac23 -2\times\sfrac{15}{13}-\sfrac23-\sfrac{53}{39}=2
\ee
This is rather remarkable and points to a definite procedure to incorporate decoupled or ``spurious'' fields. We will use these values of $h_B$ and $h_D$ to determine the conformal blocks and normalisations. In the next section we will see that this approach also works for $E_7$.

With the above data we can write down the conformal blocks and normalisation factor from Appendix C. These come out to be:
\be
\begin{split}
f^{(n)}_1(z)&=\big(z(1-z)\big)^{-\frac43}{}_2F_1\Big(n-\sfrac{17}{13},-n,;-\sfrac{2}{13};z\Big)\\[2mm]
f^{(n)}_2(z) &= {\cal N}^{(n)}\big(z(1-z)\big)^{-\frac43}z^{\frac23}\, {}_2F_1\left(n-\sfrac{2}{13},\sfrac{15}{13}-n;\sfrac{28}{13};z\right)\\[2mm]
\cN_n&=\frac{\Gamma \left(-\frac{15}{13}\right)}{\Gamma
   \left(\frac{15}{13}\right)}
   \sqrt{\frac{\Gamma \left(\frac{30}{13}-n\right) \Gamma
   \left(1+n\right)}{\Gamma
   \left(n-\frac{17}{13}\right)\Gamma
   \left(-n\right) }}
\end{split}
\ee
We see that the normalisation factor vanishes for $n=0,1,2$ confirming as expected that there is a single conformal block.

These blocks can be written in terms of elementary functions:
\be
\begin{split}
f^{(0)}_1(z)&=(z(1-z))^{-\frac{4}{3}}=(z(1-z))^{-\frac{1}{3}}\Big(\frac{1}{z}+\frac{1}{1-z}\Big)\\
f^{(1)}_1(z)&=(z(1-z))^{-\frac{4}{3}}(1-2z)=(z(1-z))^{-\frac{1}{3}}\Big(\frac{1}{z}-\frac{1}{1-z}\Big)\\
f^{(2)}_1(z)&=(z(1-z))^{-\frac{4}{3}}(1+9z-9z^2)
=(z(1-z))^{-\frac{1}{3}}\Big(\frac{1}{z}+\frac{1}{1-z}+9\Big)
\end{split}
\ee

\subsubsection{(E$_7$)$_1$}

Here we have a real primary with $h_A=\frac34$. The fusion rules are those of the SU(2)$_1$ theory, so there is just one conformal block. For E$_7$ we have:
\be
\mathbf{56}\otimes \mathbf{56} = 1\oplus \mathbf{133}\oplus\mathbf{1463}\oplus\mathbf{1539}
\ee 
so there are 4 output representations, of which the last three decouple at $k=1$. Their indices are respectively $(12, 36,660,648)$ from which their conformal dimensions are found to be:
\be
\mathbf{56}:~h=\sfrac34,\quad
\mathbf{133}:~h=\sfrac{18}{19},\quad
\mathbf{1463}:~h=\sfrac{30}{19},\quad
\mathbf{1539}:~h=\sfrac{28}{19}
\ee
From this we deduce that one should have $N=3$. The Riemann identity for the real case then implies that $h_B=\sfrac43$. But again we are in a situation where there is no single $h_B$. In its place there are three spurious fields, namely $\mathbf{133}, \mathbf{1463}, \mathbf{1539}$, whose conformal dimensions are listed above. Remarkably, the average dimension of these three primaries is seen to be $\sfrac43$. 

Thus we can use the formulae for conformal blocks of real fields with this value of $h_B$. The result, from Appendix B, is:
\be
\begin{split}
f^{(n)}_1(z)&=\big(z(1-z)\big)^{-\frac32}{}_2F_1\Big(n-\sfrac53,-n;-\sfrac13;z\Big)\\[2mm]
f^{(n)}_2(z) &= {\cal N}^{(n)} z^\frac43\big(z(1-z)\big)^{-\frac32} {}_2F_1\left(n-\sfrac13,\sfrac43-n;\sfrac73;z\right)\\[2mm]
\cN^{(n)}&=\frac{\Gamma \left(-\frac{4}{3}\right)}{\Gamma \left(\frac{4}{3}\right)}
   \sqrt{\frac{\Gamma \left(\frac{8}{3}-n\right) \Gamma
   (n+1)}{\Gamma \left(n-\frac{5}{3}\right) \Gamma
   (-n)}}
\end{split}
\ee
Again, as expected, for all allowed values of $n$ namely $n=0,1,2,3$, the normalisation factor vanishes, the second block thereby decouples and we are left with a single block. 

The blocks can be written in terms of elementary functions:
\be
\begin{split}
f^{(0)}_1(z)&=(z(1-z))^{-\frac{3}{2}}=(z(1-z))^{-\frac{1}{2}}\Big(\frac{1}{z}+\frac{1}{1-z}\Big)\\
f^{(1)}_1(z)&=(z(1-z))^{-\frac{3}{2}}(1-2z)=(z(1-z))^{-\frac{1}{2}}\Big(\frac{1}{z}-\frac{1}{1-z}\Big)\\
f^{(2)}_1(z)&=(z(1-z))^{-\frac{3}{2}}(1+2z-2z^2)
=(z(1-z))^{-\frac{1}{2}}\Big(\frac{1}{z}+\frac{1}{1-z}+2\Big)\\
f^{(3)}_1(z)&=(z(1-z))^{-\frac{3}{2}}(1+12z-42z^2+28z^3)\\
&
=(z(1-z))^{-\frac{1}{2}}\Big(\frac{1}{z}-\frac{1}{1-z}+14(1-2z)\Big)
\end{split}
\ee

Satisfyingly the answers for $E_6$ and $E_7$ reduce to elementary functions, as expected from the fact that these can be described as free scalar theories on the lattice of the corresponding Lie algebra. From the Wronskian approach, these  cases required a speculation about the way spurious primaries should be incorporated. It is likely there is a more rigorous way to understand their role. Note that in contrast, the $G_2$ and $F_4$ cases were straightforward in the Wronskian approach without additional assumptions, and (because these algebras are not simply laced) the answers are not elementary and really do require hypergeometric functions. 

\subsubsection{Non-unitary minimal model}

For completeness let us discuss the first theory listed in \cite{Mathur:1988na}, the non-unitary minimal model with $c=-\frac{22}{5}$ and a primary of dimension $h_A=-\frac15$. This case was also worked out in \cite{Mathur:1988gt}. The fusion rules are $\phi_A\times\phi_A=\II+\phi_A$. The primary is non-degenerate, so we must have $N=0$. The same result is independently confirmed by using \eref{hbsolv} to determine:
\be
h_B=-\frac15-\frac{N}{3}
\ee
We see that in order to have $h_A=h_B$ we indeed have to set $N=0$. There are two blocks, given by:
\be
\begin{split}
f_1(z)&=\big(z(1-z)\big)^{\frac25}{}_2F_1\Big(\sfrac35,\sfrac45;\sfrac65;z\Big)\\[2mm]
f_2(z) &={\cal N} \big(z(1-z)\big)^{\frac25}z^{-\frac15}\, {}_2F_1\left(\sfrac25,\sfrac35;\sfrac45;z\right)\\[2mm]
\cN&=\frac{\Gamma \left(\frac{1}{5}\right)}{\Gamma
   \left(-\frac{1}{5}\right)}
   \sqrt{\frac{\Gamma \left(\frac{1}{5}\right) \Gamma
   \left(\frac{2}{5}\right)}{\Gamma
   \left(\frac{3}{5}\right) \Gamma
   \left(\frac{4}{5}\right)}}
\end{split}
\ee

\section{Correlators of the Baby Monster CFT}
\label{babymon}

Let us now review the Baby Monster CFT. This was originally discussed in \cite{Hoehn:thesis,Hoehn:Baby8} where it was shown that the characters encode the dimensions of representations of the Baby Monster group, much in the way that the Klein $j$-invariant (with an appropriate constant added) reproduces the dimensions of representations of the Monster group. 

The Baby Monster theory was recently re-discovered in \cite{Hampapura:2016mmz} by looking for three-character RCFT's with no Kac-Moody currents. The results of that search were reported in Table 1 of that reference. The theory in the last line of the table has central charge $c=\frac{47}{2}$ and two primaries other than the identity, with dimensions $h_1=\frac{31}{16}$ and $h_2=\frac{3}{2}$ and this is the Baby Monster CFT. Its characters satisfy a differential equation constructed there and it is easy to obtain the $q$-expansion to any desired order. Concretely, to the first few orders one finds the three characters to be:
\be
\begin{split}
\tchi_0 &= q^{-\frac{47}{48}} \Big (1+ 96256 q^2 + 9646891 q^3 +\cdots\Big)\\
\tchi_1 &= q^{\frac{23}{24}} \Big(96256 +10602496 q + 420831232 q^2 +\cdots\Big)\\
\tchi_2 &=q^{\frac{25}{48}} \Big(4371 + 1143745 q+ 64680601 q^2 +\cdots\Big) 
\end{split}
\ee
which confirms, with very little work, the results originally obtained in \cite{Hoehn:thesis,Hoehn:Baby8} using Vertex Operator Algebras (VOA). 

Additionally, it was shown in \cite{Hampapura:2016mmz} that these characters satisfy a bilinear relation with the characters $\chi_i$ of the Ising model, such that the bilinear combination is the character of the $c=24$ Monster CFT:
\be
\sum_{i=0,1,2}\tchi_i(\tau)\chi_i(\tau)=j(\tau)-744
\ee
This establishes the Baby Monster CFT as a sort of generalised coset, analogous to the ones defined in \cite{Gaberdiel:2016zke}, of the Monster CFT by the Ising model. One useful consequence is that the fusion rules of the Baby Monster are the same as those of the Ising model. This fact will be used when we compute the correlation functions.

It is worth remarking that although nominally the Baby Monster is made out of 47 copies of the Ising model, it is rather difficult to explicitly construct its characters just from this observation. For example, the identity character $\tchi_0$ of the Baby Monster is not simply the 47th power of the identity character $\chi_0$ of the Ising model. This is basically because numerous spin-2 primaries contribute to this character, such as $\psi^4, \sigma^8\psi^3,\sigma^{16}\psi^2, \sigma^{24}\psi$ and $\sigma^{32}$ (this is to be understood as shorthand for the product of the corresponding fields across different copies of the Ising model). The utility of the differential equation/Wronskian approach is that we do not require any of this information to compute the characters and correlators.

Next we would like to compute correlation functions of the primaries in this theory. Since it has $c>1$ there are no Virasoro null vectors. Moreover it has no current-algebra. In the absence of these algebras and their null vectors, it is  not obvious how to approach the computation of correlators. Fortunately the method we have reviewed in the previous section enables such a computation. Let us consider the primary of dimension $\frac{31}{16}$. Since the fusion rules are known, we can be sure that the 4-point function of this primary (which we call $\phi_A$) has two conformal blocks, one corresponding to the identity and the other to the conformal family of $\phi_B$, the field with $h_B=\frac32$. We also know that $\phi_A$ primary is real and, from the characters, that it has a degeracy of 96256. Therefore we must employ the methodology of Section 3. 

The first step is to compute the value of $N$ in \eref{hbsolv}. Inserting $h_A=\frac{31}{16}$ and $h_B=\frac{3}{2}$, we find that $N=12$. This means that the parameter $n$ in \eref{degencorrtwo} takes values from 0 to 12, and there are 13 different four-point functions $G^{(n)}(z_i,\zbar_i)$ to be computed. However since the Baby Monster CFT has no Kac-Moody algebra, its chiral algebra has a minimum spin of 2. It follows that there is no first-level secondary above the identity operator. This in turn means that $G^{(n=11)}$ vanishes. Thus there are 12 correlation functions left to compute. From \eref{degsolution}, inserting the known values of $h_A$ and $N$, we find that the  two conformal blocks for each $n$ turn out to be:
\be
\begin{split}
f^{(n)}_1(z)&=\big(z(1-z)\big)^{-\frac{31}{8}}{}_2F_1\Big(-\sfrac{25}{4}+n,\sfrac{17}{4}-n;-\shalf;z\Big)\\[2mm]
f^{(n)}_2(z) &= \cN^{(n)}\big(z(1-z)\big)^{-\frac{31}{8}}z^{\frac32} {}_2F_1\left(-\sfrac{19}{4}+n,\sfrac{23}{4}-n;\sfrac52;z\right)
\end{split}
\label{babysolution}
\ee
As was the case for the Ising model, these blocks too can be expressed in terms of elementary functions. For example with $n=5$ one finds that:
\be
\begin{split}
{}_2F_1\Big(-\sfrac{5}{4},-\sfrac{3}{4};-\shalf;z\Big)&= \sfrac14\left((1+\sqrt{z})^\half + (1-\sqrt{z})^\half\right)(2-3z)\\
&\quad-\sfrac14\left((1+\sqrt{z})^\half - (1-\sqrt{z})^\half\right)\sqrt{z}\\
z^{\frac32}{}_2F_1\Big(\sfrac{1}{4},\sfrac{3}{4};\sfrac52;z\Big)&= -\sfrac25\left((1+\sqrt{z})^\half - (1-\sqrt{z})^\half\right)(2-3z)\\
&\quad+\sfrac25\left((1+\sqrt{z})^\half + (1-\sqrt{z})^\half\right)\sqrt{z}
\end{split}
\ee
The corresponding expressions in terms of elementary functions for $n=1$ are given in Appendix D. 

Under crossing, the blocks transform into each other via the matrix $M^{(n)}$ given by:
\be
\cM^{(n)}=
\begin{pmatrix}
~~\frac{\Gamma\big(-\half\big)\Gamma\big(\frac32\big)}{\Gamma\big(\frac{23}{4}-n\big)\Gamma\big(n-\frac{19}{4}\big)}
&~
\frac{1}{\cN^{(n)}}\frac{\Gamma\big(-\half\big)\Gamma\big(-\frac32\big)}{\Gamma\big(n-\frac{25}{4}\big)\Gamma\big(\frac{17}{4}-n\big)}\\[4mm]
\cN^{(n)}  \frac{\Gamma\big(\frac52\big)\Gamma\big(\frac32\big)}{\Gamma\big(\frac{29}{4}-n\big)\Gamma\big(n-\frac{13}{4}\big)} &
~~
\frac{\Gamma\big(\frac52\big)\Gamma\big(-\frac32\big)}{\Gamma\big(\frac{23}{4}-n\big)\Gamma\big(n-\frac{19}{4}\big)}
\end{pmatrix}
\ee
From unitarity of this matrix, one finds the normalisation constants:
\be
\begin{split}
 \cN^{(n)}&=\left|
\frac{\Gamma(-\frac32)}{\Gamma(\frac32) }
\sqrt\frac{\Gamma(\frac{29}{4} - n) \Gamma(n-\frac{13}{4})}{\Gamma(n-\frac{25}{4}) \Gamma(\frac{17}{4}-n)}\right|\\
&=\frac{|(17-4n)(21-4n)(25-4n)|}{24}
\end{split}
\ee

The case of the primary of dimension $\frac32$ is more non-trivial. For this primary the fusion rules, inherited from the Ising model as described above, are $\phi_A\times \phi_A=I$ and there should be just one conformal block. This means we are in the case of ``spurious'' primaries, and the analysis is more complicated. We leave this case, along with that of various coset models, for the future. 

\section{Conclusions}

We have seen that the Wronskian method, relying only on a knowledge of conformal dimensions and fusion rules, provides explicit expressions for the conformal blocks whenever a given correlator receives contributions from at most two conformal blocks. This allows us to find a second-order differential equation and solve for the correlators. In this approach, families of RCFT's are classified not by their chiral algebra but by the order of differential equation satisfied by their characters. 

It will be interesting to apply these observations to the coset models, also having small numbers of characters, discovered and studied in \cite{Naculich:1988xv,Hampapura:2015cea} and identified as novel cosets in \cite{Gaberdiel:2016zke}. The main stumbling block is to understand the role of the integer $N$ that tells us how many different tensor structures contribute to the correlator. This integer can be computed from the Riemann identity as long as there are no ``spurious'' fields. For WZW models this integer can also be computed in a straightforward way from group theory, and in all cases we examined, the two methods agree. However when there are spurious fields one has to necessarily resort to group theory and this is more tricky for coset models which are not themselves WZW models. Moreover the degeneracies of primaries in these cosets are generically rather large. We hope to report on this case in the future. 

The holomorphic bootstrap approach extends to torus correlators \cite{Mathur:1988yx,Mathur:1988jg,Mathur:1988gt, Dolan:2007eh,Gaberdiel:2008ma}. It should be possible to systematise this and obtain universal formulae for correlators which can then be evaluated for families of RCFT's just by plugging in the conformal dimensions. More interesting, and challenging, would be to understand whether this method is useful in irrational CFT contexts such as Liouville theory or logarithmic CFT \cite{Flohr:2005cm} among others. While those theories do not have finitely many characters, the method for correlators on the plane only requires sufficiently restrictive fusion rules which may exist in certain cases. 

\section*{Acknowledgements}

We would like to thank Harsha Hampapura for helpful discussions and J\"urgen Fuchs for a useful correspondence. We are grateful to Rahul Poddar for his considerable help with this revised version. The work of SM was partially supported by a J.C. Bose Fellowship, DST, Government of India. SM is grateful for the hospitality of the University of Amsterdam, the Institute for Advanced Study, Princeton, and McGill University, Montreal, where parts of this work were done. GM would like to acknowledge a DST INSPIRE Fellowship, Government of India.


\begin{appendix}

\section{Calculation of monodromy and normalisation in the non-degenerate case}

We consider the transformation of conformal blocks under the crossing $z\to 1-z$. We start by listing the relevant identities\footnote{The first relation holds only when $c-a-b$ is not an integer, which is true in all the cases we consider {\em except} SO(8), which we have treated separately.} involving hypergeometric functions (in what follows, $F(a,b;c;z)$ stands for ${}_2F_1(a,b;c;z)$):
\be
\begin{split}
F(a,b,c;1-z)&=\frac{\Gamma(c)\Gamma(c-a-b)}{\Gamma(c-a)\Gamma(c-b)}F(a,b;a+b-c+1;z)\\[2mm]
&\qquad +\frac{\Gamma(c)\Gamma(a+b-c)}{\Gamma(a)\Gamma(b)}z^{c-a-b}F(c-a,c-b;c-a-b+1;z)
\end{split}
\ee
and:
\be
F(a,b;c;z)=(1-z)^{c-a-b}F(c-a,c-b;c;z)
\ee

Rewriting the first equation with $(a,b,c)\to (c-a,c-b,c-a-b+1)$:
\be
\begin{split}
F(c-a,c-b,c-a-b+1;1-z)&=\frac{\Gamma(c-a-b+1)\Gamma(1-c)}{\Gamma(1-a)\Gamma(1-b)}F(c-a,c-b;c;z)\\[2mm]
&\quad +\frac{\Gamma(c-a-b+1)\Gamma(c-1)}{\Gamma(c-a)\Gamma(c-b)}z^{1-c}F(1-a,1-b;2-c;z)
\end{split}
\ee
And using the second equation,
\be
\begin{split}
F(c-a,c-b;c;z)&=(1-z)^{a+b-c}F(a,b;c;z)\\
F(1-a,1-b;2-c;z)&=(1-z)^{a+b-c}F(1+a-c,1+b-c;2-c;z)
\end{split}
\ee
Therefore:
\be
\begin{split}
(1-z)^{c-a-b}F(c-a,c-b;c-a-b+1;1-z)&=\frac{\Gamma(c-a-b+1)\Gamma(1-c)}{\Gamma(1-a)\Gamma(1-b)}F(a,b;c;z)\\[2mm]
&\hspace*{-4cm}+\frac{\Gamma(c-a-b+1)\Gamma(c-1)}{\Gamma(c-a)\Gamma(c-b)}z^{1-c}F(1+a-c,1+b-c;2-c;z)
\end{split}
\ee
Using the above equations and inserting the values:
\be
a=\frac13\Big(1-4h_A\Big),\quad b=-4h_A,\quad c=\frac23\Big(1-4h_A\Big)
\ee
we find:
\be
\begin{split}
&F\left(\sfrac13(1-4h_A),-4h_A;\sfrac23(1-4h_A);1-z\right) =\\[2mm]
&\quad\qquad
\frac{\Gamma(\frac23(1-4h_A))\Gamma(\frac13(1+8h_A))}{\Gamma(\frac13(1-4h_A))\Gamma(\frac23(1+2h_A))}
F\left(\sfrac13(1-4h_A),-4h_A;\sfrac23(1-4h_A);z\right)\\[2mm]
&\quad\quad
 + \frac{\Gamma(\frac23(1-4h_A))\Gamma(-\frac13(1+8h_A))}{\Gamma(\frac13(1-4h_A))\Gamma(-4h_A)}
z^{\frac13(1+8h_A)}F\left(\sfrac13(1-4h_A),\sfrac23(1+2h_A);\sfrac43(1+2h_A);z\right)\\[3mm]
&(1-z)^{\frac13(1+8h_A)}F\left(\sfrac13(1-4h_A),\sfrac23(1+2h_A);\sfrac43(1+2h_A);1-z\right) =\\[2mm]
&\quad\qquad
  \frac{\Gamma(\frac43(1+2h_A))\Gamma(\frac13(1+8h_A))}{\Gamma(\frac23(1+2h_A))\Gamma(1+4h_A)}
F\left(\sfrac13(1-4h_A),-4h_A;\sfrac23(1-4h_A);z\right)\\[2mm]
&\quad\quad +
\frac{\Gamma(\frac43(1+2h_A))\Gamma(-\frac13(1+8h_A))}{\Gamma(\frac13(1-4h_A))\Gamma(\frac23(1+2h_A))}
z^{\frac13(1+8h_A)}
F\left(\sfrac13(1-4h_A),\sfrac23(1+2h_A);\sfrac43(1+2h_A);z\right)
\end{split}
\ee

Now we define the two normalised solutions to be:
\be
\begin{split}
k_1(z)&=F\left(\sfrac13(1-4h_A),-4h_A;\sfrac23(1-4h_A);z\right)\\
k_2(z)&= \cN z^{\frac13(1+8h_A)}F\left(\sfrac13(1-4h_A),\sfrac23(1+2h_A);\sfrac43(1+2h_A);z\right)
\end{split}
\ee
where $\cN$ is the normalisation factor. Then:
\be
\begin{pmatrix} k_1(1-z)\\ k_2(1-z)\end{pmatrix}=\cM\cdot
\begin{pmatrix} k_1(z)\\ k_2(z)\end{pmatrix}
\ee
where:
\be
\cM=
\begin{pmatrix}
~~\frac{\Gamma(\frac23(1-4h_A))\Gamma(\frac13(1+8h_A))}{\Gamma(\frac13(1-4h_A))\Gamma(\frac23(1+2h_A))}
&~
\frac{1}{\cN}\frac{\Gamma(\frac23(1-4h_A))\Gamma(-\frac13(1+8h_A))}{\Gamma(\frac13(1-4h_A))\Gamma(-4h_A)}\\[4mm]
\cN  \frac{\Gamma(\frac43(1+2h_A))\Gamma(\frac13(1+8h_A))}{\Gamma(\frac23(1+2h_A))\Gamma(1+4h_A)} &
~~
\frac{\Gamma(\frac43(1+2h_A))\Gamma(-\frac13(1+8h_A))}{\Gamma(\frac13(1-4h_A))\Gamma(\frac23(1+2h_A))}
\end{pmatrix}
\ee
Crossing symmetry is achieved if the matrix $\cM$ is unitary. The condition for this is found to be:
\be
\cN^2=-\left[\frac{\Gamma(-\frac13(1+8h_A))}{\Gamma(\frac13(1+8h_A))}\right]^2
\frac{\Gamma(1+4h_A)}{\Gamma(-4h_A)}
\frac{\Gamma(\frac23(1+2h_A))}{\Gamma(\frac13(1-4h_A))}
\label{normnondeg}
\ee
Let us evaluate this for the Ising model. If we take $h_A=\frac{1}{16}$, we find $\cN^2=\frac14$. Hence $\cN=\half$, a well-known result that was used above. On the other hand, taking $h_A=\half$ gives us $\cN=0$ which reproduces the familiar result that this correlator has only one conformal block.

\section{Calculation of monodromy and normalisation for real primaries with degeneracies}

This time, the two solutions are taken to be:
\be
\begin{split}
k^{(n)}_1(z)&={}_2F_1\Big(\sfrac13(1-4h_A-N+3n),-4h_A+N-n;\sfrac23(1-4h_A)+\sfrac{N}{3};z\Big)\\[2mm]
k^{(n)}_2(z) &= {\cal N}^{(n)} z^{\frac{8h_A+1-N}{3}} {}_2F_1\left(\sfrac23(1+2h_A-N)+n,\sfrac{1-4h_A+2N}{3}-n;\sfrac43(1+2h_A)-\sfrac{N}{3};z\right)
\end{split}
\label{gensol}
\ee
and we need to compute the $\cN^{(n)}$ using crossing symmetry. From the above equations, one finds:
\be
\begin{pmatrix} k_1^{(n)}(1-z)\\ k_2^{(n)}(1-z)\end{pmatrix}=\cM^{(n)}\cdot
\begin{pmatrix} k_1^{(n)}(z)\\ k_2^{(n)}(z)\end{pmatrix}
\ee
where:
\be
\cM^{(n)}=
\begin{pmatrix}
~~\frac{\Gamma\big(\frac23(1-4h_A)+\frac{N}{3}\big)\Gamma\big(\frac13(1+8h_A)-\frac{N}{3}\big)}{\Gamma\big(\frac13(1-4h_A)+\frac{2N}{3}-n\big)\Gamma\big(\frac23(1+2h_A)-\frac{2N}{3}+n\big)}
&~
\frac{1}{\cN^{(n)}}\frac{\Gamma\big(\frac23(1-4h_A)+\frac{N}{3}\big)\Gamma\big(-\frac13(1+8h_A)+\frac{N}{3}\big)}{\Gamma\big(\frac13(1-4h_A)-\frac{N}{3}+n\big)\Gamma\big(-4h_A+N-n\big)}\\[4mm]
\cN^{(n)} \frac{\Gamma\big(\frac43(1+2h_A)-\frac{N}{3}\big)\Gamma\big(\frac13(1+8h_A)-\frac{N}{3}\big)}{\Gamma\big(\frac23(1+2h_A)+\frac{N}{3}-n\big)\Gamma\big(1+4h_A-N+n\big)} &
~~
\frac{\Gamma\big(\frac43(1+2h_A)-\frac{N}{3}\big)\Gamma\big(-\frac13(1+8h_A)+\frac{N}{3}\big)}{\Gamma\big(\frac13(1-4h_A)+\frac{2N}{3}-n\big)\Gamma\big(\frac23(1+2h_A)-\frac{2N}{3}+n\big)}
\end{pmatrix}
\label{Mmatrixreal}
\ee
Inserting $h_A=\frac{3}{5}$ and $N=4$, this agrees with Eq.(4.42) of \cite{Mathur:1988gt}.

Imposing the condition that this matrix be unitary, we find that:
\be
\begin{split}
\cN^{(n)}&=\left|\frac{\Gamma(-\frac13(1+8h_A)+\frac{N}{3})}{\Gamma(\frac13(1+8h_A)-\frac{N}{3})}
\sqrt{\frac{\Gamma(1+4h_A-N+n)}{\Gamma(-4h_A+N-n)}
\frac{\Gamma(\frac23(1+2h_A)+\frac{N}{3}-n)}{\Gamma(\frac13(1-4h_A)-\frac{N}{3}+n)}}\,\right|\\
&=\left|\frac{\Gamma(-h_B)}{\Gamma(h_B)}
\sqrt{\frac{\Gamma(-4h_A+3h_B+n)}{\Gamma(1+4h_A-3h_B-n)}
\frac{\Gamma(1+4h_A-h_B-n)}{\Gamma(-4h_A+h_B+n)}}\,\right|
\end{split}
\label{normdegreal}
\ee
The first line is written by eliminating $h_B$ in favour of $N$ while the second line retains $h_B$ and eliminates $N$. The formula is somewhat more compact in the latter version.

\section{Calculation of monodromy and normalisation for complex primaries with degeneracies}

In this case, following analogous manipulations to those done above, one finds the monodromy matrix:
\begin{equation}
\begin{pmatrix}
\frac{\Gamma\Big(\frac{1}{2}(1-8h_A+h_C+h_D+N)\Big)\Gamma\Big(\frac{1}{2}(1+8h_A-h_C-h_D-N)\Big)}{\Gamma\Big(\frac{1}{2}(1+h_C-h_D+N-2n)\Big)\Gamma\Big(\frac{1}{2}(1-h_C+h_D-N+2n)\Big)} & \frac{1}{\cN^{(n)}}\frac{\Gamma\Big(\frac{1}{2}(1-8h_A+h_C+h_D+N)\Big)\Gamma\Big(-\frac{1}{2}(1+8h_A-h_C-h_D-N)\Big)}{\Gamma\Big(-4h_A+h_D+n\Big)\Gamma\Big(-4h_A+h_C+N-n\Big)} \\
\cN^{(n)}\frac{\Gamma\Big(\frac{1}{2}(3+8h_A-h_C-h_D-N)\Big)\Gamma\Big(\frac{1}{2}(1+8h_A-h_C-h_D-N)\Big)}{\Gamma\Big(1+4h_A-h_D-n\Big)\Gamma\Big(1+4h_A-h_C-N+n\Big)} & \frac{\Gamma\Big(\frac{1}{2}(3+8h_A-h_C-h_D-N)\Big)\Gamma\Big(-\frac{1}{2}(1+8h_A-h_C-h_D-N)\Big)}{\Gamma\Big(\frac{1}{2}(1+h_C-h_D+N-2n)\Big)\Gamma\Big(\frac{1}{2}(1-h_C+h_D-N+2n)\Big)}
 \end{pmatrix}
\end{equation}
From this, the normalization constant is determined to be:
\begin{equation}
\begin{split}
\cN^{(n)} &= 
\left|\frac{\Gamma\Big(-\frac{1}{2}(1+8h_A-h_C-h_D-N)\Big)}{\Gamma\Big(\frac{1}{2}(1+8h_A-h_C-h_D-N)\Big)} \sqrt{\frac{\Gamma\Big(1+4h_A-h_D-n\Big)\Gamma\Big(1+4h_A-h_C-N+n\Big)}{\Gamma\Big(-4h_A+h_D+n\Big)\Gamma\Big(-4h_A+h_C+N-n\Big)}}\right|\\
&=\left|\frac{\Gamma(-h_B)}{\Gamma(h_B)} \sqrt{\frac{\Gamma\Big(1+4h_A-h_D-n\Big)\Gamma\Big(-4h_A+2h_B+h_D+n\Big)}{\Gamma\Big(-4h_A+h_D+n\Big)\Gamma\Big(1+4h_A-2h_B-h_D-n\Big)}}\right|
\end{split}
\label{normdegcomplex}
\end{equation}
As before, we have written the answer first in a form where $h_B$ is eliminated and then in a form where $N$ is eliminated. 

\section{Conformal blocks for Baby Monster CFT}

In this section we present the explicit form of one set of conformal blocks for the Baby Monster correlator computed in Section \ref{babymon}. They can all be written out in terms of elementary functions, much like the Ising model -- though of course, they are more complicated. There are altogether 12 cases, corresponding to $0\le n\le 12, n\ne 11$. Here we only write out the $n=1$ case. It illustrates the power of our method, which can generate a number of complicated conformal blocks from a simple, unified starting point.

\begin{equation}
\begin{split}
k_1^{(1)}(z)&=
\frac{1}{4} \sqrt{\sqrt{z}+1}~ \Big(-1664 z^{9/2}+3328
   z^{7/2}-2016 z^{5/2}+352 z^{3/2}\\
   &\qquad\quad +3328 z^5-7488 z^4+5488
   z^3-1400 z^2+69 z-\sqrt{z}+2\Big)\\
   &\quad +\frac{1}{4}
   \sqrt{1-\sqrt{z}} ~\Big(1664 z^{9/2}-3328 z^{7/2}+2016
   z^{5/2}-352 z^{3/2}\\
   &\qquad \quad +3328 z^5-7488 z^4+5488 z^3-1400 z^2+69
   z+\sqrt{z}+2\Big)\\
z^{\frac32}k_2^{(1)}(z)&=\frac{1}{4} \sqrt{\sqrt{z}+1} ~\Big(-1664 z^{9/2}+3328 z^{7/2}-2016
   z^{5/2}+352 z^{3/2}\\
   &\qquad\quad +3328 z^5-7488 z^4+5488 z^3-1400 z^2+69
   z-\sqrt{z}+2\Big)\\
   &   -\frac{1}{4} \sqrt{1-\sqrt{z}}~
   \Big(1664 z^{9/2}-3328 z^{7/2}+2016 z^{5/2}-352 z^{3/2}\\
   &\qquad\quad +3328
   z^5-7488 z^4+5488 z^3-1400 z^2+69 z+\sqrt{z}+2\Big)
\end{split}
\end{equation}
where we recall that, in general, $k_i(z)\equiv \big(z(1-z)\big)^{2h_A}f_i(z)$ and $f_i$ are the solutions of the original differential equation.

\end{appendix}


\bibliographystyle{JHEP}
\bibliography{universal}

\end{document}